\documentstyle[draft]{l-aa}
\input prepictex
\input pictex
\input postpictex
\newlength{\FigWidth}
\newlength{\FigHeight}
\def\aaa{3}
\newcommand{\dimaplota}{\rule{0.2em}{0.2em}}
\newcommand{\dimaplotb}{\rectangle <1\rc> <1\rc> }
\newcommand{\dimaplotc}{{\vpt$\bullet$}}
\newcommand{\dimaplotd}{\raisebox{-2pt}{\dima H}}
\newcommand{\dimaplote}{\raisebox{2pt}{\dima N}}
\newcommand{\dimaplotf}{\raisebox{-2pt}{\dima O}}
\newcommand{\dimaplotg}{\raisebox{2pt}{\dima M}}
\font\dima=msam5
\newlength{\rc}
\def\figskip{\vskip4ex}
\newcommand{\shifttoleft}{&&\hskip-4\arraycolsep}
\newcommand{\dd}{{\rm d}}
\newcommand{\ee}{{\rm e}}
\begin{document}
\thesaurus{03 (11.03.1; 11.05.1; 11.09.2; 11.12.2; 11.19.7)}
\title{Mergers of galaxies in clusters:
Monte Carlo simulation of mass and
angular momentum distribution}
\author{D.~S.~Krivitsky \and V.~M.~Kontorovich}
\offprints{D.~S.~Krivitsky}
\institute{Institute of Radio Astronomy, Krasnoznamennaya~4,
310002, Kharkov, Ukraine\\
e-mail: rai@ira.kharkov.ua}
\date{Received \dots; accepted \dots}
\maketitle
\markboth
 {D.~S.~Krivitsky \& V.~M.~Kontorovich: Mergers of galaxies in clusters}
 {D.~S.~Krivitsky \& V.~M.~Kontorovich: Mergers of galaxies in clusters}
\begin{abstract}
A Monte Carlo simulation of mergers in clusters of galaxies is
carried out. An ``explosive'' character of the merging process (an
analog of phase transition), suggested earlier by Cavaliere et~al.\
(\cite{ccm91}), Kontorovich et~al.\  (\cite{jetplet}), is confirmed.
In particular, a giant object similar to cD-galaxy is formed in a
comparatively short time as a result of mergers. Mass and angular
momentum distribution function for galaxies is calculated. An
intermediate asymptotics of the mass function is close to a power
law with the exponent $\alpha\approx2$. It may correspond to
recent observational data for steep faint end of the luminosity
function.  The angular momentum distribution formed by mergers is
close to Gaussian, the rms dimensionless angular momentum
$S/(GM^3R)^{1/2}$ being approximately independent of mass, which
is in accordance with observational data.
\end{abstract}

\section{Introduction}
In recent years, evolution of galaxy clusters has attracted ever
more attention. On the one hand, it has become accessible for
observations, both with ground-based and space instruments. On the
other hand, discovery of Butcher--Oemler effect testifies to an
epoch of fast evolution in clusters, associated with galaxy
interaction. Mergers of galaxies\footnote{Another possible
explanation is ``galaxy harassment'' discussed by Moore et~al.\
(\cite{moore}).\label{ftn1}} are considered to be one of the most
probable explanations for the change of colour which accompanies
appearance of ellipticals and lenticulars instead of early-type
spirals at $z\sim0.2\mbox{--}0.4$ (see, e.g., Dressler et~al.\
\cite{dres} and cited there).  Comparatively fast evolution of
clusters and groups, caused by mergers, is also confirmed by the
results of direct numerical simulation (see, e.g., Barnes
\cite{barnes}).

One of the effects associated with mergers is rapid evolution of
the galaxy mass and angular momentum distribution function
$f(M,\vec{S},t)$, related to appearance of massive galaxies.
Calculation of $f(M,\vec{S},t)$ is of great interest both by
itself and in the context of a merger model of activity suggested
by Kats \& Kontorovich (\cite{kkjetp}, \cite{kkpazh}). Given
$f(M,\vec{S},t)$, one can predict the luminosity function of
active galactic nuclei in this model.

The mass function of merging galaxies $\Phi(M,t)$ can be described
in terms of the Smoluchowski kinetic equation. As Cavaliere et~al.\
(\cite{ccm91}, \cite{ccm92}) and, independently, Kontorovich
et~al.\ (\cite{jetplet}), Kats et~al.\  (\cite{aat92}) have shown,
mergers result in an analog of phase transition (or ``explosive
evolution''), which implies fast formation of a distribution tail
(corresponding to massive galaxies) and a ``new phase'':
cD-galaxies\footnote{This transition has been known for a long
time in other applications of the Smoluchowski equation (see
Stockmayer \cite{stock}; Trubnikov \cite{trub}; Voloshchuk
\cite{vol}; Ernst \cite{fractal}).}.

The joint distribution function which takes into consideration
both mass and angular momentum can be described by a generalized
Smoluchowski equation (see Kats \& Kontorovich \cite{kkjetp},
\cite{kkobzor}, where this equation was solved in the simplest
case of a constant merger probability -- an analog of phase
transition does not take place in this variant -- and without
allowance for the orbital angular momentum). In this paper we
present the results of simulation of galaxy mass and momentum
evolution in clusters, with the orbital momentum and
more realistic
mass dependence of the merger probability taken into account.

In Sect.~\ref{sec2} we discuss Monte Carlo simulation of mergers.
In Sect.~\ref{sec3} we compare the results for $\Phi(M,t)$ with a
direct numerical solution of the Smoluchowski equation.
Section~\ref{sec4} contains discussion of the results.

\section{Joint mass and angular momentum distribution: Monte Carlo
simulation}
\label{sec2}
\subsection{Generalized Smoluchowski equation and merger probability}
The kinetic equation which describes $f(M,\vec{S},t)$ (the
generalized Smoluchowski equation) is
\begin{eqnarray}
\shifttoleft\frac{\partial f}{\partial t}
=\int W_{M\vec{S}|M_1\vec{S}_1,M_2\vec{S}_2}f_1f_2
  \,\dd M_1\,\dd M_2\,\dd^3S_1\,\dd^3S_2
\nonumber\\
\shifttoleft\qquad-2\int W_{M_1\vec{S}_1|M\vec{S},M_2\vec{S}_2}ff_2
  \,\dd M_1\,\dd M_2\,\dd^3S_1\,\dd^3S_2,
\label{1}
\end{eqnarray}
where $f_1\equiv f(M_1,\vec{S}_1,t)$ etc.
The function
$W_{M\vec{S}|M_1\vec{S}_1,M_2\vec{S}_2}$
(the kernel of the equation) is a characteristic of the
probability for the merger $M_1\vec{S}_1,M_2\vec{S}_2\to
M\vec{S}$. Taking into account mass and momentum conservation
laws, $W$~can be rewritten as
\begin{eqnarray}
\shifttoleft W_{M\vec{S}|M_1\vec{S}_1,M_2\vec{S}_2}=
U_{M\vec{S}|M_1\vec{S}_1,M_2\vec{S}_2}\nonumber\\
\shifttoleft \qquad\times\delta(M-M_1-M_2)
\delta(\vec{S}-\vec{S}_1-\vec{S}_2)
\label{2}
\end{eqnarray}
(without orbital momentum). If we take into consideration the
orbital momentum~$\vec{J}$, the second $\delta$-function should be
replaced by a function of a finite width~$\sim\overline{J}$. The
function $U$ can be calculated as $U=\overline{\sigma v}$, where
$\sigma$ is the merger cross-section, $v$~is the relative velocity
at infinity, the bar means an average over velocities.

An exact computation of collision dynamics and determining the
merger cross-section is a very complicated problem. Nevertheless,
main features of this process are known, both from analytical
consideration and numerical experiments (Roos \& Norman \cite{rn};
Aarseth \& Fall \cite{af}; Farouki \& Shapiro \cite{fs82}; Farouki
et~al.\ \cite{fsd83}; Chatterjee \cite{chat}) and enables one to
formulate conditions necessary for a merger: the galaxies must
pass at a small distance (interaction is especially intense if the
outer parts overlap) and the relative velocity must be small
enough.  Namely, we shall assume below that a merger occurs
provided that (i)~the minimal distance between the two galaxies is
less than the sum of their radii $R_1+R_2$ and (ii)~the relative
velocity at infinity is less than some limit value which is of the
order of the escape velocity $v_{\rm
g}=\sqrt{2G(M_1+M_2)/(R_1+R_2)}$, i.e., $v\le\zeta v_{\rm g}$,
$\zeta\sim1$. Taking into account gravitational focusing, we may
derive from the former condition that the impact parameter
$p_\infty\le(R_1+R_2)\sqrt{1+v_{\rm g}^2/v^2}$. Thus, the merger
cross-section is
\begin{equation}
\sigma=\cases{\pi(R_1+R_2)^2(1+v_{\rm g}^2/v^2),
&$v\le\zeta v_{\rm g}$\cr
0,&$v>\zeta v_{\rm g}$.}
\label{3}
\end{equation}
We shall assume that the galaxy peculiar velocity distribution is
Gaussian\footnote{It is essential for the further consideration
that $v_{\rm rms}$ does not depend on mass. Such a behaviour is
typical for gravitating systems which have passed through the
violent relaxation stage (Saslaw \cite{saslaw}).} with the root
mean square velocity $v_{\rm rms}$ (obviously, the relative
velocity distribution (at infinity) in this case is also Gaussian,
but the root mean square velocity is $\sqrt{2}v_{\rm rms}$). So,
\begin{eqnarray}
\shifttoleft
U=\overline{\sigma v}=\int_0^{\zeta v_{\rm g}}\pi(R_1+R_2)^2
(1+v_{\rm g}^2/v^2)v
\nonumber\\
\shifttoleft\qquad\times(4\pi/3)^{-3/2}v_{\rm rms}^{-3}
\exp(-3v^2/4v_{\rm rms}^2)4\pi v^2\,\dd v.
\label{4}
\end{eqnarray}
After integration we obtain:
\begin{eqnarray}
\shifttoleft U=4(\pi/3)^{1/2}v_{\rm rms}(R_1+R_2)^2
\nonumber\\
\shifttoleft\qquad\times\left[A+1-\ee^{-\zeta^2A}(\zeta^2A+A+1)\right],
\quad A=\frac{3v_{\rm g}^2}{4v_{\rm rms}^2}.
\label{5}
\end{eqnarray}

If the density of a galaxy $\rho_{\rm gal}$ does not depend on its
mass, then the radius $R$ and mass $M$ are related as $R\propto
M^{1/3}$. For Faber--Jackson and Tully--Fisher laws ($L\propto
V^4$), using the virial theorem ($MV^2\propto GM^2/R$) and the
mass--luminosity relation of the form $L\propto M$, we obtain
$R\propto M^{1/2}$. Below we shall take
\begin{equation}
R=CM^\beta,\qquad\mbox{where }\beta=
{\textstyle\frac13\mbox{--}\frac12}.
\label{6}
\end{equation}

Asymptotical behaviour of Eq.~(\ref{5}) is
\begin{equation}
U\approx\cases{c_2(M_1+M_2)^2,&$M\ll M_{\rm b}$\cr
c_{1+\beta}(M_1+M_2)\*(M_1^\beta+M_2^\beta),&$M\gg M_{\rm b}$,}
\label{7}
\end{equation}
the coefficients $c_2=3(3\pi)^{1/2}(\zeta^2+\zeta^4/2)G^2/v_{\rm
rms}^3$, $c_{1+\beta}=2(3\pi)^{1/2}CG/v_{\rm rms}$. Here
mass $M_{\rm b}$ corresponds to $v_{\rm rms}\sim v_{\rm g}$:
\begin{equation}
M_{\rm b}\sim(Cv_{\rm rms}^2/G)^{1/(1-\beta)}\sim
10^9\mbox{--}10^{10}\,v_7^{2/(1-\beta)}~M_\odot,
\label{8}
\end{equation}
where $v_7=v_{\rm rms}/10^7\ {\rm cm~s^{-1}}$.
We shall assume
$\zeta=1$ in the further consideration.

Note that the difference of $U$ in the small mass region from that
in Cavaliere et~al.\ \cite{ccm91} is caused by the velocity
restriction for mergers in Eq.~(\ref{3}). For $M<M_{\rm b}$
collisions without mergers are more probable than mergers
($\sigma=\pi(R_1+R_2)^2$). The latter, however, gives the
``explosive'' evolution in the small mass region too. Effects
related to collisions without mergers may be important (see
footnote~\ref{ftn1}), however, this question is beyond the scope
of this paper.

As $v_{\rm rms}$ is, in general, a function of time, the
coefficients $c_2$ and $c_{1+\beta}$ and mass $M_{\rm b}$ depends
on time too (see, e.g., Kontorovich et~al.\  \cite{jetplet}).
Below we shall assume them to be constant, neglecting
changes of velocities and masses both due to
capture of new members or evaporation of galaxies from clusters
and groups, and mergers itself, or other reasons.

In the region $M\gg M_{\rm b}$ galaxy peculiar velocities are much
less than stellar velocities in the galaxy; the relationship for
$M\ll M_{\rm b}$ is inverse. In this paper we consider
both the asymptotical regions $M\gg M_{\rm b}$ and $M\ll M_{\rm
b}$ and the intermediate case where $M_{\rm b}$ is within the
range of simulated masses.

Numerical experiments of galaxy merging (Farouki \&
Shapiro \cite{fs82}) show that the merger probability $U$ depends
both on masses and momenta, reaching a maximum when $\vec{S}_1$,
$\vec{S}_2$ and $\vec{J}$ have the same direction. Nevertheless,
this dependence is less essential than the dependence on masses
($\propto M^u$, $u=1+\beta$ for $M\gg M_{\rm b}$, which leads to
the ``explosive'' evolution). In this paper we shall use the
simple model (\ref{3}) which does not take into account the
dependence of the merger probability on the mutual orientation of
angular momenta.

It is known from the Smoluchowski equation theory (Ernst
\cite{fractal}; Voloshchuk \cite{vol}) that behaviour of the
solution essentially depends both on the homogeneity power $u$ in
the mass dependence of $U$ and on the asymptotics of $U$ for very
different masses which can be characterized by
exponents~$u_{1,2}$:
\begin{equation}
U(M_1,M_2)\approx c_u M_1^{u_1}M_2^{u_2},
\quad M_1\ll M_2, \quad u_1+u_2=u.
\end{equation}
For the case of galaxy mergers (Eq.~(\ref{7})), obviously, $u_1=0$
and $u_2=u=2$ ($M\ll M_{\rm b}$) or $1+\beta$ ($M\gg M_{\rm b}$).
If $u>1$ then an analog of phase transition takes place in the
system. For an initial distribution localized in the small mass
region, a slowly decreasing distribution tail is formed; in a
finite time the tail reaches infinite masses (in an idealized
case). The second moment of the distribution becomes infinite at
$t=t_{\rm cr}$. This phenomenon may, in principle, lead to fast
formation of massive galaxies and quasars by mergers (Kontorovich
et~al.\  \cite{jetplet}), and also to formation of cD-galaxies in
groups and clusters (Cavaliere et~al.\  \cite{ccm91}).

\subsection{Numerical simulation of mergers: methods}
Direct analytical or numerical solving the generalized
Smoluchowski equation with orbital momentum is a very complicated
problem due to the great number of variables. This difficulty can
be avoided by using numerical Monte Carlo simulation of merging
process. In this section we present such simulation. A finite
system consisting of $N$ galaxies
(referred to as ``cluster'' below, though it may also be a group)
was considered.
Pairs of these galaxies merged (with probability, proportional to
$U(M_1,M_2,\vec{S}_1,\vec{S}_2)$) until the number of the galaxies
reduced to some~$N_{\rm f}$. For each merger mass and angular
momentum conservation laws $M=M_1+M_2$,
$\vec{S}=\vec{S}_1+\vec{S}_2+\vec{J}$ were fulfilled. Distribution
of the angular momenta (intrinsic $\vec{S}_{1,2}$ and
orbital~$\vec{J}$) over directions was taken isotropic. It was
assumed that the merger probability $U$ depends only on masses
according to Eqs.\ (\ref{5}) and (\ref{7})
and does not depend on momenta.
The absolute value of the orbital momentum was computed in
accordance with Eq.~(\ref{3}), namely,
$J=\frac{M_1M_2}{M_1+M_2}vp_\infty$, $v$ being a random number
distributed in the range 0 to $v_{\rm g}$ with probability density
$f(v)\propto v^2\exp(-3v^2/4v^2_{\rm rms})$, and the impact
parameter $p_\infty$ being a random number distributed in the
range 0 to $(R_1+R_2)(1+v^2_{\rm g}/v^2)^{1/2}$ with probability
density $f(p_\infty)\propto p_\infty$.  The initial galaxy mass
distribution was chosen exponential $\Phi_0(M)\propto\ee^{-M/M_0}$
(the results are independent of the exact expression for initial
distribution, only the fact that it is localized in the region of
small masses $\sim M_0$ and decreases rapidly for large ones is
essential). Instead of the momentum~$S$, it is convenient to
introduce a dimensionless momentum
$\Lambda=\frac{S}{MR(2GM/R)^{1/2}}$, similar to Peebles' parameter
$(SE^{1/2})/(GM^{5/2})$ (see also Doroshkevich \cite{dor}).
Dimensionless momenta $\Lambda$ of the initial galaxies were
distributed uniformly in the range 0~to~1 in our simulation (as
for $\Phi_0(M)$, the exact form of the distribution is
unessential). Computations were carried out for the following
parameters: $\beta=\frac13\mbox{--}\frac12$,
$N=10^3\mbox{--}10^5$, $N_{\rm f}=(1\mbox{--}4){\cdot}10^{-1}N$.

\begin{figure}
\typeout{Drawing figure...}%
\centerline{%
\beginpicture
\setcoordinatesystem units <\FigWidth,\FigHeight>
\setplotarea x from 0 to 1, y from 0 to 1
\put {\includegraphics{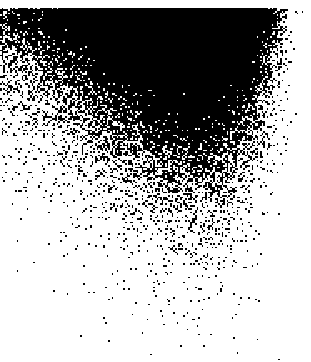}} at 0 0
\axis bottom label {$\log_{10} M$} ticks in long numbered
  withvalues $-2$ $-1$ $0$ $1$ $2$ $3$ $4$ /
  quantity 7 /
\axis left label {$\log_{10} \Lambda$} ticks in long numbered
  withvalues $-4$ $-3$ $-2$ $-1$ $0$ $1$  /
  quantity 6 /
\axis top ticks in long
  quantity 7 /
\axis right ticks in long quantity 6 /
\put {\bf a} at 0.85 0.2
\endpicture
}
\figskip
\centerline{%
\beginpicture
\setcoordinatesystem units <\FigWidth,\FigHeight>
\setplotarea x from 0 to 1, y from 0 to 1
\put {\includegraphics{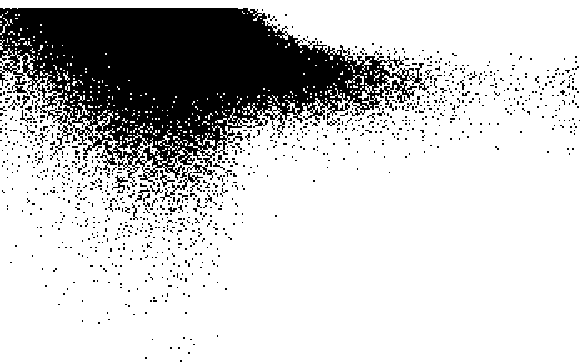}} at 0 0
\axis bottom label {$\log_{10} M$} ticks in long numbered
  withvalues $-2$ $-1$ $0$ $1$ $2$ $3$ $4$ /
  quantity 7 /
\axis left label {$\log_{10} \Lambda$} ticks in long numbered
  withvalues $-4$ $-3$ $-2$ $-1$ $0$ $1$  /
  quantity 6 /
\axis top ticks in long
  quantity 7 /
\axis right ticks in long quantity 6 /
\put {$U\propto(M_1+M_2)(M_1^{1/3}+M_2^{1/3})$} at 0.5 0.9
\put {\bf b} at 0.85 0.2
\endpicture
}
\figskip
\setlength{\unitlength}{1\FigWidth}
\centerline{%
\beginpicture
\setcoordinatesystem units <\FigWidth,\FigHeight>
\setplotarea x from 0 to 1, y from 0 to 1
\put {\includegraphics{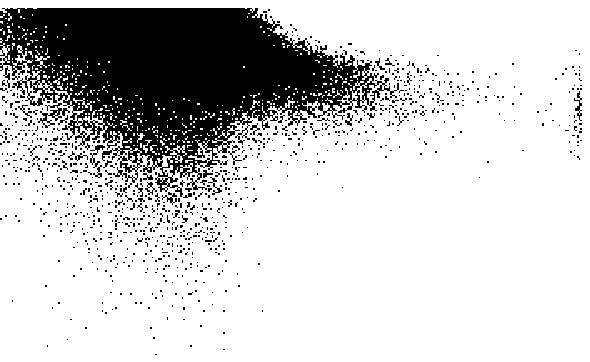}} at 0 0
\axis bottom label {$\log_{10} M$} ticks in long numbered
  withvalues $-2$ $-1$ $0$ $1$ $2$ $3$ $4$ /
  quantity 7 /
\axis left label {$\log_{10}\Lambda$} ticks in long numbered
  withvalues $-4$ $-3$ $-2$ $-1$ $0$ $1$  /
  quantity 6 /
\axis top ticks in long
  quantity 7 /
\axis right ticks in long quantity 6 /
\put {$U\propto(M_1+M_2)(M_1^{1/2}+M_2^{1/2})$} at 0.5 0.9
\put{\vector(1,1){0.06}} at 0.918 0.47
\put{cD} at 0.855 0.39
\put {\bf c} at 0.85 0.2
\endpicture
}
\caption[]{Masses $M$ and dimensionless angular momenta $\Lambda$
of simulated galaxies ($M\gg M_{\rm b}$): {\bf a}~initial, {\bf
b}~and {\bf c} formed by mergers (each dot represents one galaxy).
A distribution tail, independent of the initial conditions, is
formed due to mergers.  The right-hand part of the figure
($M\sim10^4$) corresponds to cD-galaxies. For $u=3/2$ (Fig.~{\bf
c}) separation of galaxies into the two phases, normal and cD, can
be seen better than for $u=4/3$ (Fig.~{\bf b}). Each diagram shows
$10^5$ galaxies ({\bf b} and {\bf c} -- 100 clusters of $N_{\rm
f}=1000$ galaxies each); $N_{\rm f}=10^{-1}N$, $N=10^4$.}
\label{fig1}
\typeout{End drawing.}%
\end{figure}

\begin{figure}
\typeout{Drawing figure...}%
\setlength{\unitlength}{1\FigWidth}
\centerline{%
\beginpicture
\setcoordinatesystem units <\FigWidth,\FigHeight>
\setplotarea x from 0 to 1, y from 0 to 1
\put {\includegraphics{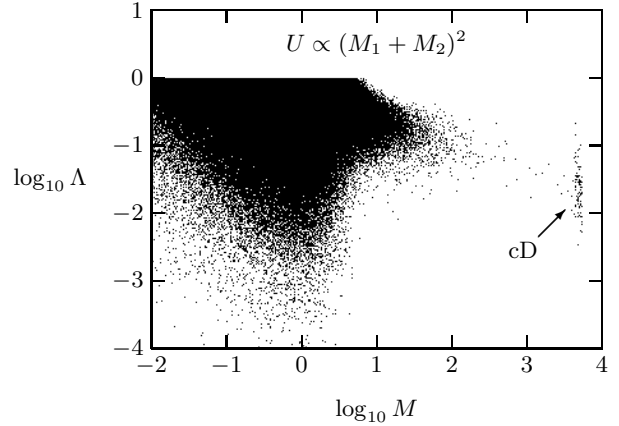}} at 0 0
\axis bottom label {$\log_{10} M$} ticks in long numbered
  withvalues $-2$ $-1$ $0$ $1$ $2$ $3$ $4$ /
  quantity 7 /
\axis left label {$\log_{10} \Lambda$} ticks in long numbered
  withvalues $-4$ $-3$ $-2$ $-1$ $0$ $1$  /
  quantity 6 /
\axis top ticks in long
  quantity 7 /
\axis right ticks in long quantity 6 /
\put {$U\propto(M_1+M_2)^2$} at 0.5 0.9
\put{\vector(1,1){0.06}} at 0.888 0.37
\put{cD} at 0.825 0.29
\endpicture
}
\caption[]{Masses $M$ and dimensionless angular momenta $\Lambda$
of simulated galaxies for the case $M\ll M_{\rm b}$. $N=10^4$,
$N_{\rm f}=4{\cdot}10^{-1}N$, 100 clusters
($4{\cdot}10^5$ galaxies).}
\label{fig1prime}
\typeout{End drawing.}%
\end{figure}

In conclusion we describe the procedure of simulation.
\begin{enumerate}
\item $N$ initial galaxies are simulated, with mass and angular
momentum distributed according to $f_0(M,\vec{S})$.
\item Two random integer numbers, $i$~and~$j$, distributed
uniformly in the range $[1,n]$ ($n$~is the current number of
galaxies) and satisfying the condition $i\ne j$, are simulated.
\item Galaxies number $i$ and $j$ merge with probability
$p=U(M_i,M_j)/U_{\rm max}$. The mass and momentum of the new
galaxy are calculated as described above. With probability $(1-p)$
the galaxies do not merge, and jump to item~2 is executed. Here
$U_{\rm max}=\max\limits_{1\le i,j\le n}U(M_i,M_j)$ depends on
time.
\item Items 2--3 are executed until the number of galaxies
$n$ becomes equal~$N_{\rm f}$.
\end{enumerate}

The algorithm used in our simulation was somewhat different from
the simplified scheme given above. The reason was that the merger
probability $p=U(M_i,M_j)/U_{\rm max}$ is very small for the
majority of galaxies and so the simulation time is very large. In
actual simulation the algorithm was modified as follows:
\begin{itemize}
\item the probability to choose the $i$-th galaxy in item~2 was
$k_i/N$ instead of $1/N$, where $k_i$ is the number of initial
galaxies which have subsequently merged into the $i$-th galaxy;
\item to compensate this change, the function
$U'(M_i,M_j)=U(M_i,M_j)/(k_ik_j)$ was used instead of $U(M_i,M_j)$
in item~3.
\end{itemize}
Obviously, these modifications do not influence the result of the
simulation. In the same time, the number of cycles reduces because
the average value of $U'/U'_{\rm max}$ is closer to~1 than the
average value of $U/U_{\rm max}$.

\subsection{Numerical simulation of mergers: results}
After some time, an analog of phase transition takes place in the
system of merging galaxies, similarly to what occurred in the work
by Cavaliere et~al.\  (\cite{ccm91}). The system divides into two
phases:  a giant galaxy which contains a major part of the total
mass and many small galaxies (in the cases $u=3/2$ and $u=2$ this
transition is more evident than for $u=4/3$).
These giant galaxies formed by mergers can
probably be
identified as cD-galaxies in the centres of groups or clusters
(cf.\ Hausman \& Ostriker \cite{ostriker}; Cavaliere
et~al.\ \cite{ccm91}).  The majority of low-mass galaxies are
those which have never merged (Fig.~\ref{fig1}). Such behaviour is
related to strong dependence of the merger probability on masses
($u=1+\beta$ or $u=2$) which, in turn, results in a steep mass
function ($\alpha\sim2$ for $u=1+\beta$, $\alpha\ga2$ for $u=2$,
see below).

\begin{figure}
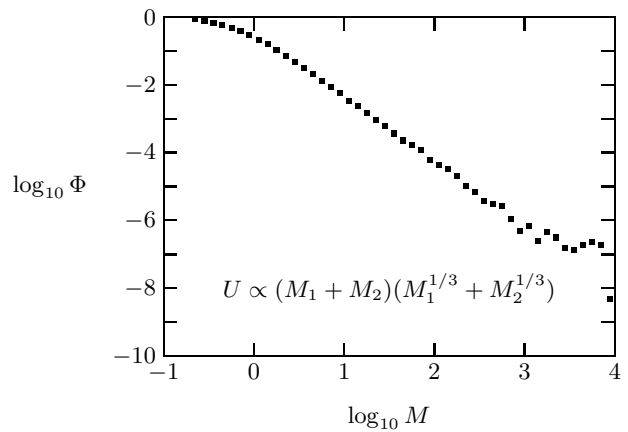

\typeout{Drawing figure...}%
\centerline{%
\beginpicture
\setcoordinatesystem units <\FigWidth,\FigHeight>
\setplotarea x from 0 to 1, y from 0 to 1
\axis bottom label {$\log_{10} M$} ticks in long numbered
  withvalues $-1$ $0$ $1$ $2$ $3$ $4$ /
  quantity 6 /
\axis left label {$\log_{10} \Phi$} ticks in long numbered
  withvalues $-10$ {} $-8$ {} $-6$ {} $-4$
  {} $-2$ {} $0$ /
  quantity 11 /
\axis top ticks in long
  quantity 6 /
\axis right ticks in long quantity 11 /
\put {$U\propto(M_1+M_2)(M_1^{1/3}+M_2^{1/3})$} at 0.5 0.2
\plotsymbolspacing50cm\setplotsymbol({\dimaplota})\plot
0.07000  0.99431
0.09000  0.98981
0.11000  0.98413
0.13000  0.97749
0.15000  0.96906
0.17000  0.95902
0.19000  0.94698
0.21000  0.93424
0.23000  0.91988
0.25000  0.90378
0.27000  0.88619
0.29000  0.86854
0.31000  0.85053
0.33000  0.83147
0.35000  0.81154
0.37000  0.79355
0.39000  0.77541
0.41000  0.75352
0.43000  0.73663
0.45000  0.71781
0.47000  0.69496
0.49000  0.67909
0.51000  0.65692
0.53000  0.63570
0.55000  0.62222
0.57000  0.60733
0.59000  0.57984
0.61000  0.56292
0.63000  0.55252
0.65000  0.53069
0.67000  0.50121
0.69000  0.48312
0.71000  0.45649
0.73000  0.44791
0.75000  0.44193
0.77000  0.40431
0.79000  0.36878
0.81000  0.38431
0.83000  0.33909
0.85000  0.36431
0.87000  0.34919
0.89000  0.31878
0.91000  0.31292
0.93000  0.32665
0.95000  0.33649
0.97000  0.32791
0.99000  0.16878
/
\endpicture
}
\caption[]{Mass function obtained by Monte Carlo simulation
($M\gg M_{\rm b}$). The values of parameters are the same as in
Fig.~\ref{fig1}. Mass is given in units~$M_0$, $\Phi(M)$~is
normalized to unity. The part of the plot near $M=10^4$
corresponds to cD-galaxies.}
\label{fig2}
\typeout{End drawing.}%
\end{figure}

\begin{figure}
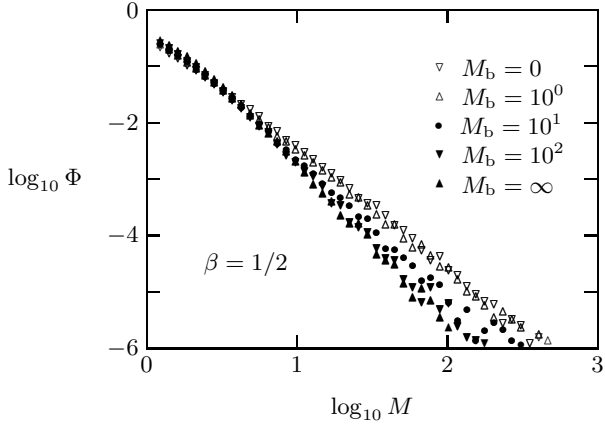

\typeout{Drawing figure...}%
\centerline{%
\beginpicture
\setcoordinatesystem units <\FigWidth,\FigHeight>
\setplotarea x from 0 to 1, y from 0 to 1
\axis bottom label {$\log_{10} M$} ticks in long numbered
  withvalues $0$ $1$ $2$ $3$ /
  quantity 4 /
\axis left label {$\log_{10} \Phi$} ticks in long numbered
  withvalues $-6$ {} $-4$
  {} $-2$ {} $0$ /
  quantity 7 /
\axis top ticks in long
  quantity 4 /
\axis right ticks in long quantity 7 /
\put {\begin{tabular}{l}
 \raisebox{4pt}{\dimaplotf}~~$M_{\rm b}=0$\\
 \raisebox{0pt}{\dimaplotg}~~$M_{\rm b}=10^0$\\
 \raisebox{2pt}{\dimaplotc}~~$M_{\rm b}=10^1$\\
 \raisebox{4pt}{\dimaplotd}~~$M_{\rm b}=10^2$\\
 \raisebox{0pt}{\dimaplote}~~$M_{\rm b}=\infty$
 \end{tabular}} at 0.8 0.65
\put {$\beta=1/2$} at 0.22 0.25
\plotsymbolspacing50cm\setplotsymbol({\dimaplotf})
\plot
0.03000  0.88946
0.05000  0.87090
0.07000  0.85608
0.09000  0.83850
0.11000  0.82097
0.13000  0.80141
0.15000  0.78079
0.17000  0.76110
0.19000  0.74093
0.21000  0.72190
0.23000  0.70649
0.25000  0.68523
0.27000  0.65658
0.29000  0.64035
0.31000  0.61662
0.33000  0.60047
0.35000  0.57998
0.37000  0.55825
0.39000  0.53578
0.41000  0.51755
0.43000  0.49603
0.45000  0.47300
0.47000  0.44319
0.49000  0.42905
0.51000  0.42184
0.53000  0.39384
0.55000  0.36456
0.57000  0.34984
0.59000  0.32449
0.61000  0.29094
0.63000  0.26195
0.65000  0.27094
0.67000  0.23019
0.69000  0.21580
0.71000  0.18497
0.73000  0.16178
0.75000  0.14001
0.77000  0.13001
0.79000  0.07398
0.81000  0.08480
0.83000  0.06850
0.85000  0.01463
0.87000  0.03398
/
\setplotsymbol({\dimaplotg})
\plot
0.03000  0.89084
0.05000  0.87524
0.07000  0.85727
0.09000  0.83697
0.11000  0.81900
0.13000  0.79802
0.15000  0.77904
0.17000  0.75405
0.19000  0.73281
0.21000  0.71334
0.23000  0.68819
0.25000  0.67075
0.27000  0.64453
0.29000  0.61939
0.31000  0.60256
0.33000  0.58093
0.35000  0.56414
0.37000  0.54201
0.39000  0.51963
0.41000  0.49887
0.43000  0.48266
0.45000  0.44810
0.47000  0.43762
0.49000  0.41500
0.51000  0.38971
0.53000  0.35935
0.55000  0.35819
0.57000  0.31884
0.59000  0.29207
0.61000  0.30130
0.63000  0.26718
0.65000  0.23580
0.67000  0.23228
0.69000  0.19793
0.71000  0.16001
0.73000  0.14563
0.75000  0.12060
0.77000  0.08398
0.79000  0.10060
0.81000  0.07850
0.83000  0.05398
0.87000  0.03398
0.89000  0.01545
/
\setplotsymbol({\dimaplotc})
\plot
0.03000  0.89935
0.05000  0.88153
0.07000  0.86399
0.09000  0.84808
0.11000  0.82466
0.13000  0.80405
0.15000  0.78264
0.17000  0.75777
0.19000  0.73536
0.21000  0.70984
0.23000  0.68683
0.25000  0.66373
0.27000  0.64010
0.29000  0.60633
0.31000  0.58622
0.33000  0.55899
0.35000  0.53990
0.37000  0.51697
0.39000  0.48836
0.41000  0.46106
0.43000  0.44442
0.45000  0.42111
0.47000  0.38841
0.49000  0.38391
0.51000  0.34232
0.53000  0.29580
0.55000  0.29228
0.57000  0.26867
0.59000  0.24497
0.61000  0.20095
0.63000  0.20806
0.65000  0.19001
0.67000  0.13398
0.69000  0.08143
0.71000  0.11398
0.73000  0.02446
0.75000  0.05143
0.77000  0.07545
0.79000  0.05579
0.81000  0.02143
0.83000  0.01143
/
\setplotsymbol({\dimaplotd})
\plot
0.03000  0.90144
0.05000  0.88602
0.07000  0.86901
0.09000  0.84890
0.11000  0.83042
0.13000  0.80709
0.15000  0.78413
0.17000  0.75864
0.19000  0.73404
0.21000  0.71011
0.23000  0.68283
0.25000  0.65776
0.27000  0.63975
0.29000  0.60249
0.31000  0.57109
0.33000  0.55043
0.35000  0.52334
0.37000  0.50698
0.39000  0.47953
0.41000  0.42984
0.43000  0.42302
0.45000  0.37367
0.47000  0.35664
0.49000  0.34207
0.51000  0.29613
0.53000  0.26531
0.55000  0.26189
0.57000  0.20480
0.59000  0.18161
0.61000  0.13463
0.63000  0.18060
0.65000  0.11463
0.67000  0.13398
0.69000  0.06528
0.71000  0.03446
0.73000  0.02446
0.75000  0.01446
/
\setplotsymbol({\dimaplote})
\plot
0.03000  0.90339
0.05000  0.88753
0.07000  0.87075
0.09000  0.85434
0.11000  0.83513
0.13000  0.81181
0.15000  0.78651
0.17000  0.76305
0.19000  0.73950
0.21000  0.71243
0.23000  0.68013
0.25000  0.65161
0.27000  0.62844
0.29000  0.60402
0.31000  0.57491
0.33000  0.54631
0.35000  0.51201
0.37000  0.47891
0.39000  0.45041
0.41000  0.42651
0.43000  0.38630
0.45000  0.36207
0.47000  0.36232
0.49000  0.32824
0.51000  0.27178
0.53000  0.25415
0.55000  0.24001
0.57000  0.18398
0.59000  0.14463
0.61000  0.17161
0.63000  0.13579
0.65000  0.08528
0.67000  0.05446
/
\endpicture
}
\caption[]{Mass function obtained by Monte Carlo simulation
for different $M_{\rm b}$ ($N_{\rm f}=10^{-1}N$).
The average slope changes with $M_{\rm b}$.}
\label{fig2prime}
\typeout{End drawing.}%
\end{figure}

The mass function formed by mergers in the case
$U\propto(M_1+M_2)\*(M_1^\beta+M_2^\beta)$ (i.e., $M\gg M_{\rm b}$)
is shown in Fig.~\ref{fig2}.  In the region $M_0\ll M\ll M_{\rm
max}$ it is close to a power law $M^{-\alpha}$, $\alpha\approx2$.
The rise near the right-hand end corresponds to cD-galaxies the
masses of which are comparable to the total masses of their
clusters. In the case $U\propto(M_1+M_2)^2$ (i.e., $M\ll M_{\rm
b}$), and in the intermediate case of a finite $M_{\rm b}$ the
mass function is steeper.

\begin{figure}
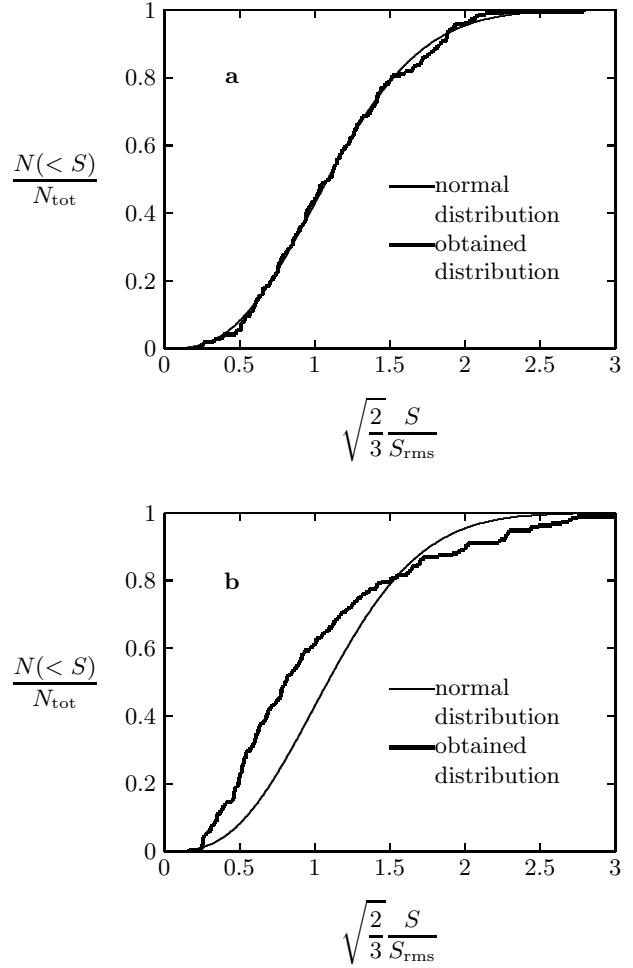

\typeout{Drawing figure...}%
\centerline{%
\beginpicture
\setcoordinatesystem units <\FigWidth,\FigHeight>
\setplotarea x from 0 to 1, y from 0 to 1
\axis bottom label {$\displaystyle\sqrt{\frac23}\frac S{S_{\rm rms}}$}
  ticks in short numbered withvalues $0$ $0.5$ $1$ $1.5$ $2$ $2.5$ $3$ /
  quantity 7 /
\axis left label
{$\displaystyle\frac{N(<S)}{N_{\rm tot}}$} ticks in short numbered
  withvalues $0$ $0.2$ $0.4$ $0.6$ $0.8$ $1$ /
  quantity 6 /
\axis top ticks in short quantity 7 /
\axis right ticks in short quantity 6 /
\put {\bf a} at 0.15 0.8
{\linethickness=\aaa\linethickness
\putrule from 0.07419 0.00000 to 0.07419 0.00461
\putrule from 0.07419 0.00461 to 0.08036 0.00461
\putrule from 0.08036 0.00461 to 0.08036 0.00922
\putrule from 0.08036 0.00922 to 0.08680 0.00922
\putrule from 0.08680 0.00922 to 0.08680 0.01382
\putrule from 0.08680 0.01382 to 0.08762 0.01382
\putrule from 0.08762 0.01382 to 0.08762 0.01843
\putrule from 0.08762 0.01843 to 0.10986 0.01843
\putrule from 0.10986 0.01843 to 0.10986 0.02304
\putrule from 0.10986 0.02304 to 0.11058 0.02304
\putrule from 0.11058 0.02304 to 0.11058 0.02765
\putrule from 0.11058 0.02765 to 0.12478 0.02765
\putrule from 0.12478 0.02765 to 0.12478 0.03226
\putrule from 0.12478 0.03226 to 0.13062 0.03226
\putrule from 0.13062 0.03226 to 0.13062 0.03687
\putrule from 0.13062 0.03687 to 0.13205 0.03687
\putrule from 0.13205 0.03687 to 0.13205 0.04147
\putrule from 0.13205 0.04147 to 0.15786 0.04147
\putrule from 0.15786 0.04147 to 0.15786 0.04608
\putrule from 0.15786 0.04608 to 0.15856 0.04608
\putrule from 0.15856 0.04608 to 0.15856 0.05069
\putrule from 0.15856 0.05069 to 0.16232 0.05069
\putrule from 0.16232 0.05069 to 0.16232 0.05530
\putrule from 0.16232 0.05530 to 0.17030 0.05530
\putrule from 0.17030 0.05530 to 0.17030 0.05991
\putrule from 0.17030 0.05991 to 0.17047 0.05991
\putrule from 0.17047 0.05991 to 0.17047 0.06452
\putrule from 0.17047 0.06452 to 0.17048 0.06452
\putrule from 0.17048 0.06452 to 0.17048 0.06912
\putrule from 0.17048 0.06912 to 0.17060 0.06912
\putrule from 0.17060 0.06912 to 0.17060 0.07373
\putrule from 0.17060 0.07373 to 0.17120 0.07373
\putrule from 0.17120 0.07373 to 0.17120 0.07834
\putrule from 0.17120 0.07834 to 0.17699 0.07834
\putrule from 0.17699 0.07834 to 0.17699 0.08295
\putrule from 0.17699 0.08295 to 0.17851 0.08295
\putrule from 0.17851 0.08295 to 0.17851 0.08756
\putrule from 0.17851 0.08756 to 0.17942 0.08756
\putrule from 0.17942 0.08756 to 0.17942 0.09217
\putrule from 0.17942 0.09217 to 0.17984 0.09217
\putrule from 0.17984 0.09217 to 0.17984 0.09677
\putrule from 0.17984 0.09677 to 0.18974 0.09677
\putrule from 0.18974 0.09677 to 0.18974 0.10138
\putrule from 0.18974 0.10138 to 0.19023 0.10138
\putrule from 0.19023 0.10138 to 0.19023 0.10599
\putrule from 0.19023 0.10599 to 0.19351 0.10599
\putrule from 0.19351 0.10599 to 0.19351 0.11060
\putrule from 0.19351 0.11060 to 0.19703 0.11060
\putrule from 0.19703 0.11060 to 0.19703 0.11521
\putrule from 0.19703 0.11521 to 0.19704 0.11521
\putrule from 0.19704 0.11521 to 0.19704 0.11982
\putrule from 0.19704 0.11982 to 0.19918 0.11982
\putrule from 0.19918 0.11982 to 0.19918 0.12442
\putrule from 0.19918 0.12442 to 0.20074 0.12442
\putrule from 0.20074 0.12442 to 0.20074 0.12903
\putrule from 0.20074 0.12903 to 0.20155 0.12903
\putrule from 0.20155 0.12903 to 0.20155 0.13364
\putrule from 0.20155 0.13364 to 0.20247 0.13364
\putrule from 0.20247 0.13364 to 0.20247 0.13825
\putrule from 0.20247 0.13825 to 0.20432 0.13825
\putrule from 0.20432 0.13825 to 0.20432 0.14286
\putrule from 0.20432 0.14286 to 0.20640 0.14286
\putrule from 0.20640 0.14286 to 0.20640 0.14747
\putrule from 0.20640 0.14747 to 0.20880 0.14747
\putrule from 0.20880 0.14747 to 0.20880 0.15207
\putrule from 0.20880 0.15207 to 0.20975 0.15207
\putrule from 0.20975 0.15207 to 0.20975 0.15668
\putrule from 0.20975 0.15668 to 0.21587 0.15668
\putrule from 0.21587 0.15668 to 0.21587 0.16129
\putrule from 0.21587 0.16129 to 0.21769 0.16129
\putrule from 0.21769 0.16129 to 0.21769 0.16590
\putrule from 0.21769 0.16590 to 0.21843 0.16590
\putrule from 0.21843 0.16590 to 0.21843 0.17051
\putrule from 0.21843 0.17051 to 0.21889 0.17051
\putrule from 0.21889 0.17051 to 0.21889 0.17512
\putrule from 0.21889 0.17512 to 0.21918 0.17512
\putrule from 0.21918 0.17512 to 0.21918 0.17972
\putrule from 0.21918 0.17972 to 0.22513 0.17972
\putrule from 0.22513 0.17972 to 0.22513 0.18433
\putrule from 0.22513 0.18433 to 0.23244 0.18433
\putrule from 0.23244 0.18433 to 0.23244 0.18894
\putrule from 0.23244 0.18894 to 0.23373 0.18894
\putrule from 0.23373 0.18894 to 0.23373 0.19355
\putrule from 0.23373 0.19355 to 0.23399 0.19355
\putrule from 0.23399 0.19355 to 0.23399 0.19816
\putrule from 0.23399 0.19816 to 0.23701 0.19816
\putrule from 0.23701 0.19816 to 0.23701 0.20276
\putrule from 0.23701 0.20276 to 0.23829 0.20276
\putrule from 0.23829 0.20276 to 0.23829 0.20737
\putrule from 0.23829 0.20737 to 0.24297 0.20737
\putrule from 0.24297 0.20737 to 0.24297 0.21198
\putrule from 0.24297 0.21198 to 0.24423 0.21198
\putrule from 0.24423 0.21198 to 0.24423 0.21659
\putrule from 0.24423 0.21659 to 0.24693 0.21659
\putrule from 0.24693 0.21659 to 0.24693 0.22120
\putrule from 0.24693 0.22120 to 0.24808 0.22120
\putrule from 0.24808 0.22120 to 0.24808 0.22581
\putrule from 0.24808 0.22581 to 0.25162 0.22581
\putrule from 0.25162 0.22581 to 0.25162 0.23041
\putrule from 0.25162 0.23041 to 0.25202 0.23041
\putrule from 0.25202 0.23041 to 0.25202 0.23502
\putrule from 0.25202 0.23502 to 0.25263 0.23502
\putrule from 0.25263 0.23502 to 0.25263 0.23963
\putrule from 0.25263 0.23963 to 0.25337 0.23963
\putrule from 0.25337 0.23963 to 0.25337 0.24424
\putrule from 0.25337 0.24424 to 0.25363 0.24424
\putrule from 0.25363 0.24424 to 0.25363 0.24885
\putrule from 0.25363 0.24885 to 0.25497 0.24885
\putrule from 0.25497 0.24885 to 0.25497 0.25346
\putrule from 0.25497 0.25346 to 0.25679 0.25346
\putrule from 0.25679 0.25346 to 0.25679 0.25806
\putrule from 0.25679 0.25806 to 0.25804 0.25806
\putrule from 0.25804 0.25806 to 0.25804 0.26267
\putrule from 0.25804 0.26267 to 0.25851 0.26267
\putrule from 0.25851 0.26267 to 0.25851 0.26728
\putrule from 0.25851 0.26728 to 0.26157 0.26728
\putrule from 0.26157 0.26728 to 0.26157 0.27189
\putrule from 0.26157 0.27189 to 0.26249 0.27189
\putrule from 0.26249 0.27189 to 0.26249 0.27650
\putrule from 0.26249 0.27650 to 0.26577 0.27650
\putrule from 0.26577 0.27650 to 0.26577 0.28111
\putrule from 0.26577 0.28111 to 0.27217 0.28111
\putrule from 0.27217 0.28111 to 0.27217 0.28571
\putrule from 0.27217 0.28571 to 0.27257 0.28571
\putrule from 0.27257 0.28571 to 0.27257 0.29032
\putrule from 0.27257 0.29032 to 0.27371 0.29032
\putrule from 0.27371 0.29032 to 0.27371 0.29493
\putrule from 0.27371 0.29493 to 0.27653 0.29493
\putrule from 0.27653 0.29493 to 0.27653 0.29954
\putrule from 0.27653 0.29954 to 0.28121 0.29954
\putrule from 0.28121 0.29954 to 0.28121 0.30415
\putrule from 0.28121 0.30415 to 0.28412 0.30415
\putrule from 0.28412 0.30415 to 0.28412 0.30876
\putrule from 0.28412 0.30876 to 0.28470 0.30876
\putrule from 0.28470 0.30876 to 0.28470 0.31336
\putrule from 0.28470 0.31336 to 0.28532 0.31336
\putrule from 0.28532 0.31336 to 0.28532 0.31797
\putrule from 0.28532 0.31797 to 0.28610 0.31797
\putrule from 0.28610 0.31797 to 0.28610 0.32258
\putrule from 0.28610 0.32258 to 0.28757 0.32258
\putrule from 0.28757 0.32258 to 0.28757 0.32719
\putrule from 0.28757 0.32719 to 0.28967 0.32719
\putrule from 0.28967 0.32719 to 0.28967 0.33180
\putrule from 0.28967 0.33180 to 0.29253 0.33180
\putrule from 0.29253 0.33180 to 0.29253 0.33641
\putrule from 0.29253 0.33641 to 0.29295 0.33641
\putrule from 0.29295 0.33641 to 0.29295 0.34101
\putrule from 0.29295 0.34101 to 0.29529 0.34101
\putrule from 0.29529 0.34101 to 0.29529 0.34562
\putrule from 0.29529 0.34562 to 0.29954 0.34562
\putrule from 0.29954 0.34562 to 0.29954 0.35023
\putrule from 0.29954 0.35023 to 0.30012 0.35023
\putrule from 0.30012 0.35023 to 0.30012 0.35484
\putrule from 0.30012 0.35484 to 0.30157 0.35484
\putrule from 0.30157 0.35484 to 0.30157 0.35945
\putrule from 0.30157 0.35945 to 0.30328 0.35945
\putrule from 0.30328 0.35945 to 0.30328 0.36406
\putrule from 0.30328 0.36406 to 0.30722 0.36406
\putrule from 0.30722 0.36406 to 0.30722 0.36866
\putrule from 0.30722 0.36866 to 0.31010 0.36866
\putrule from 0.31010 0.36866 to 0.31010 0.37327
\putrule from 0.31010 0.37327 to 0.31213 0.37327
\putrule from 0.31213 0.37327 to 0.31213 0.37788
\putrule from 0.31213 0.37788 to 0.31325 0.37788
\putrule from 0.31325 0.37788 to 0.31325 0.38249
\putrule from 0.31325 0.38249 to 0.31457 0.38249
\putrule from 0.31457 0.38249 to 0.31457 0.38710
\putrule from 0.31457 0.38710 to 0.31458 0.38710
\putrule from 0.31458 0.38710 to 0.31458 0.39171
\putrule from 0.31458 0.39171 to 0.31480 0.39171
\putrule from 0.31480 0.39171 to 0.31480 0.39631
\putrule from 0.31480 0.39631 to 0.31521 0.39631
\putrule from 0.31521 0.39631 to 0.31521 0.40092
\putrule from 0.31521 0.40092 to 0.31732 0.40092
\putrule from 0.31732 0.40092 to 0.31732 0.40553
\putrule from 0.31732 0.40553 to 0.31783 0.40553
\putrule from 0.31783 0.40553 to 0.31783 0.41014
\putrule from 0.31783 0.41014 to 0.31809 0.41014
\putrule from 0.31809 0.41014 to 0.31809 0.41475
\putrule from 0.31809 0.41475 to 0.32090 0.41475
\putrule from 0.32090 0.41475 to 0.32090 0.41935
\putrule from 0.32090 0.41935 to 0.32214 0.41935
\putrule from 0.32214 0.41935 to 0.32214 0.42396
\putrule from 0.32214 0.42396 to 0.32449 0.42396
\putrule from 0.32449 0.42396 to 0.32449 0.42857
\putrule from 0.32449 0.42857 to 0.32759 0.42857
\putrule from 0.32759 0.42857 to 0.32759 0.43318
\putrule from 0.32759 0.43318 to 0.33049 0.43318
\putrule from 0.33049 0.43318 to 0.33049 0.43779
\putrule from 0.33049 0.43779 to 0.33421 0.43779
\putrule from 0.33421 0.43779 to 0.33421 0.44240
\putrule from 0.33421 0.44240 to 0.33529 0.44240
\putrule from 0.33529 0.44240 to 0.33529 0.44700
\putrule from 0.33529 0.44700 to 0.33883 0.44700
\putrule from 0.33883 0.44700 to 0.33883 0.45161
\putrule from 0.33883 0.45161 to 0.33952 0.45161
\putrule from 0.33952 0.45161 to 0.33952 0.45622
\putrule from 0.33952 0.45622 to 0.34066 0.45622
\putrule from 0.34066 0.45622 to 0.34066 0.46083
\putrule from 0.34066 0.46083 to 0.34258 0.46083
\putrule from 0.34258 0.46083 to 0.34258 0.46544
\putrule from 0.34258 0.46544 to 0.34392 0.46544
\putrule from 0.34392 0.46544 to 0.34392 0.47005
\putrule from 0.34392 0.47005 to 0.34470 0.47005
\putrule from 0.34470 0.47005 to 0.34470 0.47465
\putrule from 0.34470 0.47465 to 0.34529 0.47465
\putrule from 0.34529 0.47465 to 0.34529 0.47926
\putrule from 0.34529 0.47926 to 0.34561 0.47926
\putrule from 0.34561 0.47926 to 0.34561 0.48387
\putrule from 0.34561 0.48387 to 0.34919 0.48387
\putrule from 0.34919 0.48387 to 0.34919 0.48848
\putrule from 0.34919 0.48848 to 0.35581 0.48848
\putrule from 0.35581 0.48848 to 0.35581 0.49309
\putrule from 0.35581 0.49309 to 0.36016 0.49309
\putrule from 0.36016 0.49309 to 0.36016 0.49770
\putrule from 0.36016 0.49770 to 0.36567 0.49770
\putrule from 0.36567 0.49770 to 0.36567 0.50230
\putrule from 0.36567 0.50230 to 0.36748 0.50230
\putrule from 0.36748 0.50230 to 0.36748 0.50691
\putrule from 0.36748 0.50691 to 0.37215 0.50691
\putrule from 0.37215 0.50691 to 0.37215 0.51152
\putrule from 0.37215 0.51152 to 0.37239 0.51152
\putrule from 0.37239 0.51152 to 0.37239 0.51613
\putrule from 0.37239 0.51613 to 0.37294 0.51613
\putrule from 0.37294 0.51613 to 0.37294 0.52074
\putrule from 0.37294 0.52074 to 0.37437 0.52074
\putrule from 0.37437 0.52074 to 0.37437 0.52535
\putrule from 0.37437 0.52535 to 0.37470 0.52535
\putrule from 0.37470 0.52535 to 0.37470 0.52995
\putrule from 0.37470 0.52995 to 0.37527 0.52995
\putrule from 0.37527 0.52995 to 0.37527 0.53456
\putrule from 0.37527 0.53456 to 0.37695 0.53456
\putrule from 0.37695 0.53456 to 0.37695 0.53917
\putrule from 0.37695 0.53917 to 0.37895 0.53917
\putrule from 0.37895 0.53917 to 0.37895 0.54378
\putrule from 0.37895 0.54378 to 0.38218 0.54378
\putrule from 0.38218 0.54378 to 0.38218 0.54839
\putrule from 0.38218 0.54839 to 0.38219 0.54839
\putrule from 0.38219 0.54839 to 0.38219 0.55300
\putrule from 0.38219 0.55300 to 0.38268 0.55300
\putrule from 0.38268 0.55300 to 0.38268 0.55760
\putrule from 0.38268 0.55760 to 0.38290 0.55760
\putrule from 0.38290 0.55760 to 0.38290 0.56221
\putrule from 0.38290 0.56221 to 0.38915 0.56221
\putrule from 0.38915 0.56221 to 0.38915 0.56682
\putrule from 0.38915 0.56682 to 0.38956 0.56682
\putrule from 0.38956 0.56682 to 0.38956 0.57143
\putrule from 0.38956 0.57143 to 0.39029 0.57143
\putrule from 0.39029 0.57143 to 0.39029 0.57604
\putrule from 0.39029 0.57604 to 0.39303 0.57604
\putrule from 0.39303 0.57604 to 0.39303 0.58065
\putrule from 0.39303 0.58065 to 0.39437 0.58065
\putrule from 0.39437 0.58065 to 0.39437 0.58525
\putrule from 0.39437 0.58525 to 0.39505 0.58525
\putrule from 0.39505 0.58525 to 0.39505 0.58986
\putrule from 0.39505 0.58986 to 0.40010 0.58986
\putrule from 0.40010 0.58986 to 0.40010 0.59447
\putrule from 0.40010 0.59447 to 0.40124 0.59447
\putrule from 0.40124 0.59447 to 0.40124 0.59908
\putrule from 0.40124 0.59908 to 0.40620 0.59908
\putrule from 0.40620 0.59908 to 0.40620 0.60369
\putrule from 0.40620 0.60369 to 0.40853 0.60369
\putrule from 0.40853 0.60369 to 0.40853 0.60829
\putrule from 0.40853 0.60829 to 0.41078 0.60829
\putrule from 0.41078 0.60829 to 0.41078 0.61290
\putrule from 0.41078 0.61290 to 0.41125 0.61290
\putrule from 0.41125 0.61290 to 0.41125 0.61751
\putrule from 0.41125 0.61751 to 0.41255 0.61751
\putrule from 0.41255 0.61751 to 0.41255 0.62212
\putrule from 0.41255 0.62212 to 0.41512 0.62212
\putrule from 0.41512 0.62212 to 0.41512 0.62673
\putrule from 0.41512 0.62673 to 0.41832 0.62673
\putrule from 0.41832 0.62673 to 0.41832 0.63134
\putrule from 0.41832 0.63134 to 0.41959 0.63134
\putrule from 0.41959 0.63134 to 0.41959 0.63594
\putrule from 0.41959 0.63594 to 0.42151 0.63594
\putrule from 0.42151 0.63594 to 0.42151 0.64055
\putrule from 0.42151 0.64055 to 0.42288 0.64055
\putrule from 0.42288 0.64055 to 0.42288 0.64516
\putrule from 0.42288 0.64516 to 0.42307 0.64516
\putrule from 0.42307 0.64516 to 0.42307 0.64977
\putrule from 0.42307 0.64977 to 0.42735 0.64977
\putrule from 0.42735 0.64977 to 0.42735 0.65438
\putrule from 0.42735 0.65438 to 0.42766 0.65438
\putrule from 0.42766 0.65438 to 0.42766 0.65899
\putrule from 0.42766 0.65899 to 0.42823 0.65899
\putrule from 0.42823 0.65899 to 0.42823 0.66359
\putrule from 0.42823 0.66359 to 0.43192 0.66359
\putrule from 0.43192 0.66359 to 0.43192 0.66820
\putrule from 0.43192 0.66820 to 0.43439 0.66820
\putrule from 0.43439 0.66820 to 0.43439 0.67281
\putrule from 0.43439 0.67281 to 0.43517 0.67281
\putrule from 0.43517 0.67281 to 0.43517 0.67742
\putrule from 0.43517 0.67742 to 0.43731 0.67742
\putrule from 0.43731 0.67742 to 0.43731 0.68203
\putrule from 0.43731 0.68203 to 0.43904 0.68203
\putrule from 0.43904 0.68203 to 0.43904 0.68664
\putrule from 0.43904 0.68664 to 0.45054 0.68664
\putrule from 0.45054 0.68664 to 0.45054 0.69124
\putrule from 0.45054 0.69124 to 0.45547 0.69124
\putrule from 0.45547 0.69124 to 0.45547 0.69585
\putrule from 0.45547 0.69585 to 0.45596 0.69585
\putrule from 0.45596 0.69585 to 0.45596 0.70046
\putrule from 0.45596 0.70046 to 0.45770 0.70046
\putrule from 0.45770 0.70046 to 0.45770 0.70507
\putrule from 0.45770 0.70507 to 0.45996 0.70507
\putrule from 0.45996 0.70507 to 0.45996 0.70968
\putrule from 0.45996 0.70968 to 0.46121 0.70968
\putrule from 0.46121 0.70968 to 0.46121 0.71429
\putrule from 0.46121 0.71429 to 0.46420 0.71429
\putrule from 0.46420 0.71429 to 0.46420 0.71889
\putrule from 0.46420 0.71889 to 0.46760 0.71889
\putrule from 0.46760 0.71889 to 0.46760 0.72350
\putrule from 0.46760 0.72350 to 0.46896 0.72350
\putrule from 0.46896 0.72350 to 0.46896 0.72811
\putrule from 0.46896 0.72811 to 0.47065 0.72811
\putrule from 0.47065 0.72811 to 0.47065 0.73272
\putrule from 0.47065 0.73272 to 0.47107 0.73272
\putrule from 0.47107 0.73272 to 0.47107 0.73733
\putrule from 0.47107 0.73733 to 0.47138 0.73733
\putrule from 0.47138 0.73733 to 0.47138 0.74194
\putrule from 0.47138 0.74194 to 0.47259 0.74194
\putrule from 0.47259 0.74194 to 0.47259 0.74654
\putrule from 0.47259 0.74654 to 0.47330 0.74654
\putrule from 0.47330 0.74654 to 0.47330 0.75115
\putrule from 0.47330 0.75115 to 0.47520 0.75115
\putrule from 0.47520 0.75115 to 0.47520 0.75576
\putrule from 0.47520 0.75576 to 0.47881 0.75576
\putrule from 0.47881 0.75576 to 0.47881 0.76037
\putrule from 0.47881 0.76037 to 0.47972 0.76037
\putrule from 0.47972 0.76037 to 0.47972 0.76498
\putrule from 0.47972 0.76498 to 0.48013 0.76498
\putrule from 0.48013 0.76498 to 0.48013 0.76959
\putrule from 0.48013 0.76959 to 0.48505 0.76959
\putrule from 0.48505 0.76959 to 0.48505 0.77419
\putrule from 0.48505 0.77419 to 0.48897 0.77419
\putrule from 0.48897 0.77419 to 0.48897 0.77880
\putrule from 0.48897 0.77880 to 0.48955 0.77880
\putrule from 0.48955 0.77880 to 0.48955 0.78341
\putrule from 0.48955 0.78341 to 0.49769 0.78341
\putrule from 0.49769 0.78341 to 0.49769 0.78802
\putrule from 0.49769 0.78802 to 0.49799 0.78802
\putrule from 0.49799 0.78802 to 0.49799 0.79263
\putrule from 0.49799 0.79263 to 0.50172 0.79263
\putrule from 0.50172 0.79263 to 0.50172 0.79724
\putrule from 0.50172 0.79724 to 0.50490 0.79724
\putrule from 0.50490 0.79724 to 0.50490 0.80184
\putrule from 0.50490 0.80184 to 0.50590 0.80184
\putrule from 0.50590 0.80184 to 0.50590 0.80645
\putrule from 0.50590 0.80645 to 0.52228 0.80645
\putrule from 0.52228 0.80645 to 0.52228 0.81106
\putrule from 0.52228 0.81106 to 0.53124 0.81106
\putrule from 0.53124 0.81106 to 0.53124 0.81567
\putrule from 0.53124 0.81567 to 0.53498 0.81567
\putrule from 0.53498 0.81567 to 0.53498 0.82028
\putrule from 0.53498 0.82028 to 0.54961 0.82028
\putrule from 0.54961 0.82028 to 0.54961 0.82488
\putrule from 0.54961 0.82488 to 0.55234 0.82488
\putrule from 0.55234 0.82488 to 0.55234 0.82949
\putrule from 0.55234 0.82949 to 0.55318 0.82949
\putrule from 0.55318 0.82949 to 0.55318 0.83410
\putrule from 0.55318 0.83410 to 0.55907 0.83410
\putrule from 0.55907 0.83410 to 0.55907 0.83871
\putrule from 0.55907 0.83871 to 0.56821 0.83871
\putrule from 0.56821 0.83871 to 0.56821 0.84332
\putrule from 0.56821 0.84332 to 0.56958 0.84332
\putrule from 0.56958 0.84332 to 0.56958 0.84793
\putrule from 0.56958 0.84793 to 0.57113 0.84793
\putrule from 0.57113 0.84793 to 0.57113 0.85253
\putrule from 0.57113 0.85253 to 0.57391 0.85253
\putrule from 0.57391 0.85253 to 0.57391 0.85714
\putrule from 0.57391 0.85714 to 0.57492 0.85714
\putrule from 0.57492 0.85714 to 0.57492 0.86175
\putrule from 0.57492 0.86175 to 0.58221 0.86175
\putrule from 0.58221 0.86175 to 0.58221 0.86636
\putrule from 0.58221 0.86636 to 0.58555 0.86636
\putrule from 0.58555 0.86636 to 0.58555 0.87097
\putrule from 0.58555 0.87097 to 0.58946 0.87097
\putrule from 0.58946 0.87097 to 0.58946 0.87558
\putrule from 0.58946 0.87558 to 0.59431 0.87558
\putrule from 0.59431 0.87558 to 0.59431 0.88018
\putrule from 0.59431 0.88018 to 0.59932 0.88018
\putrule from 0.59932 0.88018 to 0.59932 0.88479
\putrule from 0.59932 0.88479 to 0.60126 0.88479
\putrule from 0.60126 0.88479 to 0.60126 0.88940
\putrule from 0.60126 0.88940 to 0.60729 0.88940
\putrule from 0.60729 0.88940 to 0.60729 0.89401
\putrule from 0.60729 0.89401 to 0.60887 0.89401
\putrule from 0.60887 0.89401 to 0.60887 0.89862
\putrule from 0.60887 0.89862 to 0.61310 0.89862
\putrule from 0.61310 0.89862 to 0.61310 0.90323
\putrule from 0.61310 0.90323 to 0.61429 0.90323
\putrule from 0.61429 0.90323 to 0.61429 0.90783
\putrule from 0.61429 0.90783 to 0.62450 0.90783
\putrule from 0.62450 0.90783 to 0.62450 0.91244
\putrule from 0.62450 0.91244 to 0.62522 0.91244
\putrule from 0.62522 0.91244 to 0.62522 0.91705
\putrule from 0.62522 0.91705 to 0.62621 0.91705
\putrule from 0.62621 0.91705 to 0.62621 0.92166
\putrule from 0.62621 0.92166 to 0.62625 0.92166
\putrule from 0.62625 0.92166 to 0.62625 0.92627
\putrule from 0.62625 0.92627 to 0.62740 0.92627
\putrule from 0.62740 0.92627 to 0.62740 0.93088
\putrule from 0.62740 0.93088 to 0.62900 0.93088
\putrule from 0.62900 0.93088 to 0.62900 0.93548
\putrule from 0.62900 0.93548 to 0.62973 0.93548
\putrule from 0.62973 0.93548 to 0.62973 0.94009
\putrule from 0.62973 0.94009 to 0.63822 0.94009
\putrule from 0.63822 0.94009 to 0.63822 0.94470
\putrule from 0.63822 0.94470 to 0.63894 0.94470
\putrule from 0.63894 0.94470 to 0.63894 0.94931
\putrule from 0.63894 0.94931 to 0.64481 0.94931
\putrule from 0.64481 0.94931 to 0.64481 0.95392
\putrule from 0.64481 0.95392 to 0.64532 0.95392
\putrule from 0.64532 0.95392 to 0.64532 0.95853
\putrule from 0.64532 0.95853 to 0.66530 0.95853
\putrule from 0.66530 0.95853 to 0.66530 0.96313
\putrule from 0.66530 0.96313 to 0.67589 0.96313
\putrule from 0.67589 0.96313 to 0.67589 0.96774
\putrule from 0.67589 0.96774 to 0.67940 0.96774
\putrule from 0.67940 0.96774 to 0.67940 0.97235
\putrule from 0.67940 0.97235 to 0.68917 0.97235
\putrule from 0.68917 0.97235 to 0.68917 0.97696
\putrule from 0.68917 0.97696 to 0.68923 0.97696
\putrule from 0.68923 0.97696 to 0.68923 0.98157
\putrule from 0.68923 0.98157 to 0.69671 0.98157
\putrule from 0.69671 0.98157 to 0.69671 0.98618
\putrule from 0.69671 0.98618 to 0.71176 0.98618
\putrule from 0.71176 0.98618 to 0.71176 0.99078
\putrule from 0.71176 0.99078 to 0.75076 0.99078
\putrule from 0.75076 0.99078 to 0.75076 0.99539
\putrule from 0.75076 0.99539 to 0.92986 0.99539
\putrule from 0.92986 0.99539 to 0.92986 1.00000
}%
\plot
 0.00000 0.00000
 0.00667 0.00001
 0.01333 0.00005
 0.02000 0.00016
 0.02667 0.00038
 0.03333 0.00075
 0.04000 0.00129
 0.04667 0.00204
 0.05333 0.00303
 0.06000 0.00430
 0.06667 0.00588
 0.07333 0.00778
 0.08000 0.01005
 0.08667 0.01270
 0.09333 0.01576
 0.10000 0.01925
 0.10667 0.02319
 0.11333 0.02760
 0.12000 0.03249
 0.12667 0.03788
 0.13333 0.04378
 0.14000 0.05019
 0.14667 0.05713
 0.15333 0.06459
 0.16000 0.07258
 0.16667 0.08111
 0.17333 0.09016
 0.18000 0.09973
 0.18667 0.10982
 0.19333 0.12042
 0.20000 0.13151
 0.20667 0.14309
 0.21333 0.15513
 0.22000 0.16763
 0.22667 0.18056
 0.23333 0.19391
 0.24000 0.20765
 0.24667 0.22177
 0.25333 0.23623
 0.26000 0.25102
 0.26667 0.26611
 0.27333 0.28148
 0.28000 0.29709
 0.28667 0.31292
 0.29333 0.32895
 0.30000 0.34514
 0.30667 0.36146
 0.31333 0.37790
 0.32000 0.39442
 0.32667 0.41099
 0.33333 0.42759
 0.34000 0.44419
 0.34667 0.46077
 0.35333 0.47729
 0.36000 0.49373
 0.36667 0.51008
 0.37333 0.52630
 0.38000 0.54237
 0.38667 0.55828
 0.39333 0.57399
 0.40000 0.58950
 0.40667 0.60478
 0.41333 0.61982
 0.42000 0.63461
 0.42667 0.64912
 0.43333 0.66334
 0.44000 0.67726
 0.44667 0.69088
 0.45333 0.70417
 0.46000 0.71714
 0.46667 0.72977
 0.47333 0.74206
 0.48000 0.75400
 0.48667 0.76559
 0.49333 0.77683
 0.50000 0.78771
 0.50667 0.79824
 0.51333 0.80840
 0.52000 0.81822
 0.52667 0.82767
 0.53333 0.83678
 0.54000 0.84554
 0.54667 0.85395
 0.55333 0.86203
 0.56000 0.86977
 0.56667 0.87718
 0.57333 0.88427
 0.58000 0.89104
 0.58667 0.89751
 0.59333 0.90367
 0.60000 0.90955
 0.60667 0.91513
 0.61333 0.92044
 0.62000 0.92548
 0.62667 0.93026
 0.63333 0.93479
 0.64000 0.93908
 0.64667 0.94314
 0.65333 0.94697
 0.66000 0.95058
 0.66667 0.95399
 0.67333 0.95720
 0.68000 0.96022
 0.68667 0.96305
 0.69333 0.96572
 0.70000 0.96822
 0.70667 0.97056
 0.71333 0.97275
 0.72000 0.97480
 0.72667 0.97672
 0.73333 0.97851
 0.74000 0.98018
 0.74667 0.98173
 0.75333 0.98318
 0.76000 0.98452
 0.76667 0.98577
 0.77333 0.98693
 0.78000 0.98801
 0.78667 0.98900
 0.79333 0.98993
 0.80000 0.99078
 0.80667 0.99157
 0.81333 0.99229
 0.82000 0.99296
 0.82667 0.99358
 0.83333 0.99415
 0.84000 0.99467
 0.84667 0.99515
 0.85333 0.99559
 0.86000 0.99599
 0.86667 0.99636
 0.87333 0.99670
 0.88000 0.99701
 0.88667 0.99729
 0.89333 0.99755
 0.90000 0.99779
 0.90667 0.99800
 0.91333 0.99820
 0.92000 0.99837
 0.92667 0.99853
 0.93333 0.99868
 0.94000 0.99881
 0.94667 0.99893
 0.95333 0.99904
 0.96000 0.99914
 0.96667 0.99923
 0.97333 0.99931
 0.98000 0.99938
 0.98667 0.99945
 0.99333 0.99951
 1.00000 0.99956
/
{\small
\put
{\begin{tabular}{l}
normal\\distribution\\
obtained\\distribution
\end{tabular}} [l]
at 0.6 0.35
\putrule <0pt,1.5\baselineskip> from 0.5 0.35 to 0.6 0.35
\linethickness=\aaa\linethickness
\putrule <0pt,-0.5\baselineskip> from 0.5 0.35 to 0.6 0.35 }
\endpicture
}
\figskip
\centerline{%
\beginpicture
\setcoordinatesystem units <\FigWidth,\FigHeight>
\setplotarea x from 0 to 1, y from 0 to 1
\axis bottom label {$\displaystyle\sqrt{\frac23}\frac S{S_{\rm rms}}$}
  ticks in short numbered withvalues $0$ $0.5$ $1$ $1.5$ $2$ $2.5$ $3$ /
  quantity 7 /
\axis left label
{$\displaystyle\frac{N(<S)}{N_{\rm tot}}$} ticks in short numbered
  withvalues $0$ $0.2$ $0.4$ $0.6$ $0.8$ $1$ /
  quantity 6 /
\axis top ticks in short quantity 7 /
\axis right ticks in short quantity 6 /
\put {\bf b} at 0.15 0.8
{\linethickness=\aaa\linethickness
\putrule from 0.05584 0.00000 to 0.05584 0.00521
\putrule from 0.05584 0.00521 to 0.08136 0.00521
\putrule from 0.08136 0.00521 to 0.08136 0.01042
\putrule from 0.08136 0.01042 to 0.08319 0.01042
\putrule from 0.08319 0.01042 to 0.08319 0.01563
\putrule from 0.08319 0.01563 to 0.08343 0.01563
\putrule from 0.08343 0.01563 to 0.08343 0.02083
\putrule from 0.08343 0.02083 to 0.08516 0.02083
\putrule from 0.08516 0.02083 to 0.08516 0.02604
\putrule from 0.08516 0.02604 to 0.08557 0.02604
\putrule from 0.08557 0.02604 to 0.08557 0.03125
\putrule from 0.08557 0.03125 to 0.08566 0.03125
\putrule from 0.08566 0.03125 to 0.08566 0.03646
\putrule from 0.08566 0.03646 to 0.08580 0.03646
\putrule from 0.08580 0.03646 to 0.08580 0.04167
\putrule from 0.08580 0.04167 to 0.08836 0.04167
\putrule from 0.08836 0.04167 to 0.08836 0.04688
\putrule from 0.08836 0.04688 to 0.09161 0.04688
\putrule from 0.09161 0.04688 to 0.09161 0.05208
\putrule from 0.09161 0.05208 to 0.09633 0.05208
\putrule from 0.09633 0.05208 to 0.09633 0.05729
\putrule from 0.09633 0.05729 to 0.10033 0.05729
\putrule from 0.10033 0.05729 to 0.10033 0.06250
\putrule from 0.10033 0.06250 to 0.10179 0.06250
\putrule from 0.10179 0.06250 to 0.10179 0.06771
\putrule from 0.10179 0.06771 to 0.10404 0.06771
\putrule from 0.10404 0.06771 to 0.10404 0.07292
\putrule from 0.10404 0.07292 to 0.10608 0.07292
\putrule from 0.10608 0.07292 to 0.10608 0.07813
\putrule from 0.10608 0.07813 to 0.11065 0.07813
\putrule from 0.11065 0.07813 to 0.11065 0.08333
\putrule from 0.11065 0.08333 to 0.11289 0.08333
\putrule from 0.11289 0.08333 to 0.11289 0.08854
\putrule from 0.11289 0.08854 to 0.11588 0.08854
\putrule from 0.11588 0.08854 to 0.11588 0.09375
\putrule from 0.11588 0.09375 to 0.11606 0.09375
\putrule from 0.11606 0.09375 to 0.11606 0.09896
\putrule from 0.11606 0.09896 to 0.11624 0.09896
\putrule from 0.11624 0.09896 to 0.11624 0.10417
\putrule from 0.11624 0.10417 to 0.11625 0.10417
\putrule from 0.11625 0.10417 to 0.11625 0.10938
\putrule from 0.11625 0.10938 to 0.12385 0.10938
\putrule from 0.12385 0.10938 to 0.12385 0.11458
\putrule from 0.12385 0.11458 to 0.12391 0.11458
\putrule from 0.12391 0.11458 to 0.12391 0.11979
\putrule from 0.12391 0.11979 to 0.12828 0.11979
\putrule from 0.12828 0.11979 to 0.12828 0.12500
\putrule from 0.12828 0.12500 to 0.12953 0.12500
\putrule from 0.12953 0.12500 to 0.12953 0.13021
\putrule from 0.12953 0.13021 to 0.13048 0.13021
\putrule from 0.13048 0.13021 to 0.13048 0.13542
\putrule from 0.13048 0.13542 to 0.13546 0.13542
\putrule from 0.13546 0.13542 to 0.13546 0.14063
\putrule from 0.13546 0.14063 to 0.13849 0.14063
\putrule from 0.13849 0.14063 to 0.13849 0.14583
\putrule from 0.13849 0.14583 to 0.14838 0.14583
\putrule from 0.14838 0.14583 to 0.14838 0.15104
\putrule from 0.14838 0.15104 to 0.15277 0.15104
\putrule from 0.15277 0.15104 to 0.15277 0.15625
\putrule from 0.15277 0.15625 to 0.15444 0.15625
\putrule from 0.15444 0.15625 to 0.15444 0.16146
\putrule from 0.15444 0.16146 to 0.15449 0.16146
\putrule from 0.15449 0.16146 to 0.15449 0.16667
\putrule from 0.15449 0.16667 to 0.15524 0.16667
\putrule from 0.15524 0.16667 to 0.15524 0.17188
\putrule from 0.15524 0.17188 to 0.15526 0.17188
\putrule from 0.15526 0.17188 to 0.15526 0.17708
\putrule from 0.15526 0.17708 to 0.15670 0.17708
\putrule from 0.15670 0.17708 to 0.15670 0.18229
\putrule from 0.15670 0.18229 to 0.15737 0.18229
\putrule from 0.15737 0.18229 to 0.15737 0.18750
\putrule from 0.15737 0.18750 to 0.15985 0.18750
\putrule from 0.15985 0.18750 to 0.15985 0.19271
\putrule from 0.15985 0.19271 to 0.16323 0.19271
\putrule from 0.16323 0.19271 to 0.16323 0.19792
\putrule from 0.16323 0.19792 to 0.16455 0.19792
\putrule from 0.16455 0.19792 to 0.16455 0.20313
\putrule from 0.16455 0.20313 to 0.16486 0.20313
\putrule from 0.16486 0.20313 to 0.16486 0.20833
\putrule from 0.16486 0.20833 to 0.16621 0.20833
\putrule from 0.16621 0.20833 to 0.16621 0.21354
\putrule from 0.16621 0.21354 to 0.16635 0.21354
\putrule from 0.16635 0.21354 to 0.16635 0.21875
\putrule from 0.16635 0.21875 to 0.16710 0.21875
\putrule from 0.16710 0.21875 to 0.16710 0.22396
\putrule from 0.16710 0.22396 to 0.16798 0.22396
\putrule from 0.16798 0.22396 to 0.16798 0.22917
\putrule from 0.16798 0.22917 to 0.16970 0.22917
\putrule from 0.16970 0.22917 to 0.16970 0.23438
\putrule from 0.16970 0.23438 to 0.17235 0.23438
\putrule from 0.17235 0.23438 to 0.17235 0.23958
\putrule from 0.17235 0.23958 to 0.17239 0.23958
\putrule from 0.17239 0.23958 to 0.17239 0.24479
\putrule from 0.17239 0.24479 to 0.17248 0.24479
\putrule from 0.17248 0.24479 to 0.17248 0.25000
\putrule from 0.17248 0.25000 to 0.17344 0.25000
\putrule from 0.17344 0.25000 to 0.17344 0.25521
\putrule from 0.17344 0.25521 to 0.17452 0.25521
\putrule from 0.17452 0.25521 to 0.17452 0.26042
\putrule from 0.17452 0.26042 to 0.17667 0.26042
\putrule from 0.17667 0.26042 to 0.17667 0.26563
\putrule from 0.17667 0.26563 to 0.17682 0.26563
\putrule from 0.17682 0.26563 to 0.17682 0.27083
\putrule from 0.17682 0.27083 to 0.17693 0.27083
\putrule from 0.17693 0.27083 to 0.17693 0.27604
\putrule from 0.17693 0.27604 to 0.17978 0.27604
\putrule from 0.17978 0.27604 to 0.17978 0.28125
\putrule from 0.17978 0.28125 to 0.18139 0.28125
\putrule from 0.18139 0.28125 to 0.18139 0.28646
\putrule from 0.18139 0.28646 to 0.18193 0.28646
\putrule from 0.18193 0.28646 to 0.18193 0.29167
\putrule from 0.18193 0.29167 to 0.18230 0.29167
\putrule from 0.18230 0.29167 to 0.18230 0.29688
\putrule from 0.18230 0.29688 to 0.19048 0.29688
\putrule from 0.19048 0.29688 to 0.19048 0.30208
\putrule from 0.19048 0.30208 to 0.19135 0.30208
\putrule from 0.19135 0.30208 to 0.19135 0.30729
\putrule from 0.19135 0.30729 to 0.19611 0.30729
\putrule from 0.19611 0.30729 to 0.19611 0.31250
\putrule from 0.19611 0.31250 to 0.19628 0.31250
\putrule from 0.19628 0.31250 to 0.19628 0.31771
\putrule from 0.19628 0.31771 to 0.19986 0.31771
\putrule from 0.19986 0.31771 to 0.19986 0.32292
\putrule from 0.19986 0.32292 to 0.20121 0.32292
\putrule from 0.20121 0.32292 to 0.20121 0.32813
\putrule from 0.20121 0.32813 to 0.20159 0.32813
\putrule from 0.20159 0.32813 to 0.20159 0.33333
\putrule from 0.20159 0.33333 to 0.20263 0.33333
\putrule from 0.20263 0.33333 to 0.20263 0.33854
\putrule from 0.20263 0.33854 to 0.20278 0.33854
\putrule from 0.20278 0.33854 to 0.20278 0.34375
\putrule from 0.20278 0.34375 to 0.20389 0.34375
\putrule from 0.20389 0.34375 to 0.20389 0.34896
\putrule from 0.20389 0.34896 to 0.20792 0.34896
\putrule from 0.20792 0.34896 to 0.20792 0.35417
\putrule from 0.20792 0.35417 to 0.20889 0.35417
\putrule from 0.20889 0.35417 to 0.20889 0.35938
\putrule from 0.20889 0.35938 to 0.20976 0.35938
\putrule from 0.20976 0.35938 to 0.20976 0.36458
\putrule from 0.20976 0.36458 to 0.21017 0.36458
\putrule from 0.21017 0.36458 to 0.21017 0.36979
\putrule from 0.21017 0.36979 to 0.21221 0.36979
\putrule from 0.21221 0.36979 to 0.21221 0.37500
\putrule from 0.21221 0.37500 to 0.21476 0.37500
\putrule from 0.21476 0.37500 to 0.21476 0.38021
\putrule from 0.21476 0.38021 to 0.22164 0.38021
\putrule from 0.22164 0.38021 to 0.22164 0.38542
\putrule from 0.22164 0.38542 to 0.22375 0.38542
\putrule from 0.22375 0.38542 to 0.22375 0.39063
\putrule from 0.22375 0.39063 to 0.22563 0.39063
\putrule from 0.22563 0.39063 to 0.22563 0.39583
\putrule from 0.22563 0.39583 to 0.22666 0.39583
\putrule from 0.22666 0.39583 to 0.22666 0.40104
\putrule from 0.22666 0.40104 to 0.22805 0.40104
\putrule from 0.22805 0.40104 to 0.22805 0.40625
\putrule from 0.22805 0.40625 to 0.22931 0.40625
\putrule from 0.22931 0.40625 to 0.22931 0.41146
\putrule from 0.22931 0.41146 to 0.23030 0.41146
\putrule from 0.23030 0.41146 to 0.23030 0.41667
\putrule from 0.23030 0.41667 to 0.23107 0.41667
\putrule from 0.23107 0.41667 to 0.23107 0.42188
\putrule from 0.23107 0.42188 to 0.23696 0.42188
\putrule from 0.23696 0.42188 to 0.23696 0.42708
\putrule from 0.23696 0.42708 to 0.23953 0.42708
\putrule from 0.23953 0.42708 to 0.23953 0.43229
\putrule from 0.23953 0.43229 to 0.24313 0.43229
\putrule from 0.24313 0.43229 to 0.24313 0.43750
\putrule from 0.24313 0.43750 to 0.24752 0.43750
\putrule from 0.24752 0.43750 to 0.24752 0.44271
\putrule from 0.24752 0.44271 to 0.24886 0.44271
\putrule from 0.24886 0.44271 to 0.24886 0.44792
\putrule from 0.24886 0.44792 to 0.25555 0.44792
\putrule from 0.25555 0.44792 to 0.25555 0.45313
\putrule from 0.25555 0.45313 to 0.25594 0.45313
\putrule from 0.25594 0.45313 to 0.25594 0.45833
\putrule from 0.25594 0.45833 to 0.25823 0.45833
\putrule from 0.25823 0.45833 to 0.25823 0.46354
\putrule from 0.25823 0.46354 to 0.25900 0.46354
\putrule from 0.25900 0.46354 to 0.25900 0.46875
\putrule from 0.25900 0.46875 to 0.25962 0.46875
\putrule from 0.25962 0.46875 to 0.25962 0.47396
\putrule from 0.25962 0.47396 to 0.25978 0.47396
\putrule from 0.25978 0.47396 to 0.25978 0.47917
\putrule from 0.25978 0.47917 to 0.26134 0.47917
\putrule from 0.26134 0.47917 to 0.26134 0.48438
\putrule from 0.26134 0.48438 to 0.26308 0.48438
\putrule from 0.26308 0.48438 to 0.26308 0.48958
\putrule from 0.26308 0.48958 to 0.26671 0.48958
\putrule from 0.26671 0.48958 to 0.26671 0.49479
\putrule from 0.26671 0.49479 to 0.26732 0.49479
\putrule from 0.26732 0.49479 to 0.26732 0.50000
\putrule from 0.26732 0.50000 to 0.26935 0.50000
\putrule from 0.26935 0.50000 to 0.26935 0.50521
\putrule from 0.26935 0.50521 to 0.27157 0.50521
\putrule from 0.27157 0.50521 to 0.27157 0.51042
\putrule from 0.27157 0.51042 to 0.27186 0.51042
\putrule from 0.27186 0.51042 to 0.27186 0.51563
\putrule from 0.27186 0.51563 to 0.27516 0.51563
\putrule from 0.27516 0.51563 to 0.27516 0.52083
\putrule from 0.27516 0.52083 to 0.28001 0.52083
\putrule from 0.28001 0.52083 to 0.28001 0.52604
\putrule from 0.28001 0.52604 to 0.28209 0.52604
\putrule from 0.28209 0.52604 to 0.28209 0.53125
\putrule from 0.28209 0.53125 to 0.28589 0.53125
\putrule from 0.28589 0.53125 to 0.28589 0.53646
\putrule from 0.28589 0.53646 to 0.29058 0.53646
\putrule from 0.29058 0.53646 to 0.29058 0.54167
\putrule from 0.29058 0.54167 to 0.29143 0.54167
\putrule from 0.29143 0.54167 to 0.29143 0.54688
\putrule from 0.29143 0.54688 to 0.29641 0.54688
\putrule from 0.29641 0.54688 to 0.29641 0.55208
\putrule from 0.29641 0.55208 to 0.29807 0.55208
\putrule from 0.29807 0.55208 to 0.29807 0.55729
\putrule from 0.29807 0.55729 to 0.30126 0.55729
\putrule from 0.30126 0.55729 to 0.30126 0.56250
\putrule from 0.30126 0.56250 to 0.30274 0.56250
\putrule from 0.30274 0.56250 to 0.30274 0.56771
\putrule from 0.30274 0.56771 to 0.30377 0.56771
\putrule from 0.30377 0.56771 to 0.30377 0.57292
\putrule from 0.30377 0.57292 to 0.30564 0.57292
\putrule from 0.30564 0.57292 to 0.30564 0.57813
\putrule from 0.30564 0.57813 to 0.30592 0.57813
\putrule from 0.30592 0.57813 to 0.30592 0.58333
\putrule from 0.30592 0.58333 to 0.31021 0.58333
\putrule from 0.31021 0.58333 to 0.31021 0.58854
\putrule from 0.31021 0.58854 to 0.31196 0.58854
\putrule from 0.31196 0.58854 to 0.31196 0.59375
\putrule from 0.31196 0.59375 to 0.32117 0.59375
\putrule from 0.32117 0.59375 to 0.32117 0.59896
\putrule from 0.32117 0.59896 to 0.32428 0.59896
\putrule from 0.32428 0.59896 to 0.32428 0.60417
\putrule from 0.32428 0.60417 to 0.33015 0.60417
\putrule from 0.33015 0.60417 to 0.33015 0.60938
\putrule from 0.33015 0.60938 to 0.33293 0.60938
\putrule from 0.33293 0.60938 to 0.33293 0.61458
\putrule from 0.33293 0.61458 to 0.33655 0.61458
\putrule from 0.33655 0.61458 to 0.33655 0.61979
\putrule from 0.33655 0.61979 to 0.33708 0.61979
\putrule from 0.33708 0.61979 to 0.33708 0.62500
\putrule from 0.33708 0.62500 to 0.33727 0.62500
\putrule from 0.33727 0.62500 to 0.33727 0.63021
\putrule from 0.33727 0.63021 to 0.34654 0.63021
\putrule from 0.34654 0.63021 to 0.34654 0.63542
\putrule from 0.34654 0.63542 to 0.35035 0.63542
\putrule from 0.35035 0.63542 to 0.35035 0.64063
\putrule from 0.35035 0.64063 to 0.35201 0.64063
\putrule from 0.35201 0.64063 to 0.35201 0.64583
\putrule from 0.35201 0.64583 to 0.35647 0.64583
\putrule from 0.35647 0.64583 to 0.35647 0.65104
\putrule from 0.35647 0.65104 to 0.36484 0.65104
\putrule from 0.36484 0.65104 to 0.36484 0.65625
\putrule from 0.36484 0.65625 to 0.36512 0.65625
\putrule from 0.36512 0.65625 to 0.36512 0.66146
\putrule from 0.36512 0.66146 to 0.37174 0.66146
\putrule from 0.37174 0.66146 to 0.37174 0.66667
\putrule from 0.37174 0.66667 to 0.37329 0.66667
\putrule from 0.37329 0.66667 to 0.37329 0.67188
\putrule from 0.37329 0.67188 to 0.37598 0.67188
\putrule from 0.37598 0.67188 to 0.37598 0.67708
\putrule from 0.37598 0.67708 to 0.37672 0.67708
\putrule from 0.37672 0.67708 to 0.37672 0.68229
\putrule from 0.37672 0.68229 to 0.38131 0.68229
\putrule from 0.38131 0.68229 to 0.38131 0.68750
\putrule from 0.38131 0.68750 to 0.38658 0.68750
\putrule from 0.38658 0.68750 to 0.38658 0.69271
\putrule from 0.38658 0.69271 to 0.38735 0.69271
\putrule from 0.38735 0.69271 to 0.38735 0.69792
\putrule from 0.38735 0.69792 to 0.39103 0.69792
\putrule from 0.39103 0.69792 to 0.39103 0.70313
\putrule from 0.39103 0.70313 to 0.39576 0.70313
\putrule from 0.39576 0.70313 to 0.39576 0.70833
\putrule from 0.39576 0.70833 to 0.40260 0.70833
\putrule from 0.40260 0.70833 to 0.40260 0.71354
\putrule from 0.40260 0.71354 to 0.40476 0.71354
\putrule from 0.40476 0.71354 to 0.40476 0.71875
\putrule from 0.40476 0.71875 to 0.41217 0.71875
\putrule from 0.41217 0.71875 to 0.41217 0.72396
\putrule from 0.41217 0.72396 to 0.41637 0.72396
\putrule from 0.41637 0.72396 to 0.41637 0.72917
\putrule from 0.41637 0.72917 to 0.41903 0.72917
\putrule from 0.41903 0.72917 to 0.41903 0.73438
\putrule from 0.41903 0.73438 to 0.42322 0.73438
\putrule from 0.42322 0.73438 to 0.42322 0.73958
\putrule from 0.42322 0.73958 to 0.42696 0.73958
\putrule from 0.42696 0.73958 to 0.42696 0.74479
\putrule from 0.42696 0.74479 to 0.42897 0.74479
\putrule from 0.42897 0.74479 to 0.42897 0.75000
\putrule from 0.42897 0.75000 to 0.43644 0.75000
\putrule from 0.43644 0.75000 to 0.43644 0.75521
\putrule from 0.43644 0.75521 to 0.43989 0.75521
\putrule from 0.43989 0.75521 to 0.43989 0.76042
\putrule from 0.43989 0.76042 to 0.44660 0.76042
\putrule from 0.44660 0.76042 to 0.44660 0.76563
\putrule from 0.44660 0.76563 to 0.45393 0.76563
\putrule from 0.45393 0.76563 to 0.45393 0.77083
\putrule from 0.45393 0.77083 to 0.45563 0.77083
\putrule from 0.45563 0.77083 to 0.45563 0.77604
\putrule from 0.45563 0.77604 to 0.46945 0.77604
\putrule from 0.46945 0.77604 to 0.46945 0.78125
\putrule from 0.46945 0.78125 to 0.47127 0.78125
\putrule from 0.47127 0.78125 to 0.47127 0.78646
\putrule from 0.47127 0.78646 to 0.47231 0.78646
\putrule from 0.47231 0.78646 to 0.47231 0.79167
\putrule from 0.47231 0.79167 to 0.48145 0.79167
\putrule from 0.48145 0.79167 to 0.48145 0.79688
\putrule from 0.48145 0.79688 to 0.50298 0.79688
\putrule from 0.50298 0.79688 to 0.50298 0.80208
\putrule from 0.50298 0.80208 to 0.50354 0.80208
\putrule from 0.50354 0.80208 to 0.50354 0.80729
\putrule from 0.50354 0.80729 to 0.51545 0.80729
\putrule from 0.51545 0.80729 to 0.51545 0.81250
\putrule from 0.51545 0.81250 to 0.51579 0.81250
\putrule from 0.51579 0.81250 to 0.51579 0.81771
\putrule from 0.51579 0.81771 to 0.53822 0.81771
\putrule from 0.53822 0.81771 to 0.53822 0.82292
\putrule from 0.53822 0.82292 to 0.54433 0.82292
\putrule from 0.54433 0.82292 to 0.54433 0.82813
\putrule from 0.54433 0.82813 to 0.54552 0.82813
\putrule from 0.54552 0.82813 to 0.54552 0.83333
\putrule from 0.54552 0.83333 to 0.54986 0.83333
\putrule from 0.54986 0.83333 to 0.54986 0.83854
\putrule from 0.54986 0.83854 to 0.55057 0.83854
\putrule from 0.55057 0.83854 to 0.55057 0.84375
\putrule from 0.55057 0.84375 to 0.56361 0.84375
\putrule from 0.56361 0.84375 to 0.56361 0.84896
\putrule from 0.56361 0.84896 to 0.56721 0.84896
\putrule from 0.56721 0.84896 to 0.56721 0.85417
\putrule from 0.56721 0.85417 to 0.56760 0.85417
\putrule from 0.56760 0.85417 to 0.56760 0.85938
\putrule from 0.56760 0.85938 to 0.56938 0.85938
\putrule from 0.56938 0.85938 to 0.56938 0.86458
\putrule from 0.56938 0.86458 to 0.57565 0.86458
\putrule from 0.57565 0.86458 to 0.57565 0.86979
\putrule from 0.57565 0.86979 to 0.60812 0.86979
\putrule from 0.60812 0.86979 to 0.60812 0.87500
\putrule from 0.60812 0.87500 to 0.63926 0.87500
\putrule from 0.63926 0.87500 to 0.63926 0.88021
\putrule from 0.63926 0.88021 to 0.64978 0.88021
\putrule from 0.64978 0.88021 to 0.64978 0.88542
\putrule from 0.64978 0.88542 to 0.65568 0.88542
\putrule from 0.65568 0.88542 to 0.65568 0.89063
\putrule from 0.65568 0.89063 to 0.66538 0.89063
\putrule from 0.66538 0.89063 to 0.66538 0.89583
\putrule from 0.66538 0.89583 to 0.66919 0.89583
\putrule from 0.66919 0.89583 to 0.66919 0.90104
\putrule from 0.66919 0.90104 to 0.67010 0.90104
\putrule from 0.67010 0.90104 to 0.67010 0.90625
\putrule from 0.67010 0.90625 to 0.67580 0.90625
\putrule from 0.67580 0.90625 to 0.67580 0.91146
\putrule from 0.67580 0.91146 to 0.74114 0.91146
\putrule from 0.74114 0.91146 to 0.74114 0.91667
\putrule from 0.74114 0.91667 to 0.74596 0.91667
\putrule from 0.74596 0.91667 to 0.74596 0.92188
\putrule from 0.74596 0.92188 to 0.75627 0.92188
\putrule from 0.75627 0.92188 to 0.75627 0.92708
\putrule from 0.75627 0.92708 to 0.75845 0.92708
\putrule from 0.75845 0.92708 to 0.75845 0.93229
\putrule from 0.75845 0.93229 to 0.75895 0.93229
\putrule from 0.75895 0.93229 to 0.75895 0.93750
\putrule from 0.75895 0.93750 to 0.76158 0.93750
\putrule from 0.76158 0.93750 to 0.76158 0.94271
\putrule from 0.76158 0.94271 to 0.76439 0.94271
\putrule from 0.76439 0.94271 to 0.76439 0.94792
\putrule from 0.76439 0.94792 to 0.81186 0.94792
\putrule from 0.81186 0.94792 to 0.81186 0.95313
\putrule from 0.81186 0.95313 to 0.81222 0.95313
\putrule from 0.81222 0.95313 to 0.81222 0.95833
\putrule from 0.81222 0.95833 to 0.82933 0.95833
\putrule from 0.82933 0.95833 to 0.82933 0.96354
\putrule from 0.82933 0.96354 to 0.85830 0.96354
\putrule from 0.85830 0.96354 to 0.85830 0.96875
\putrule from 0.85830 0.96875 to 0.88078 0.96875
\putrule from 0.88078 0.96875 to 0.88078 0.97396
\putrule from 0.88078 0.97396 to 0.89545 0.97396
\putrule from 0.89545 0.97396 to 0.89545 0.97917
\putrule from 0.89545 0.97917 to 0.90215 0.97917
\putrule from 0.90215 0.97917 to 0.90215 0.98438
\putrule from 0.90215 0.98438 to 0.91631 0.98438
\putrule from 0.91631 0.98438 to 0.91631 0.98958
\putrule from 0.91631 0.98958 to 1.00000 0.98958
}%
\plot
 0.00000 0.00000
 0.00667 0.00001
 0.01333 0.00005
 0.02000 0.00016
 0.02667 0.00038
 0.03333 0.00075
 0.04000 0.00129
 0.04667 0.00204
 0.05333 0.00303
 0.06000 0.00430
 0.06667 0.00588
 0.07333 0.00778
 0.08000 0.01005
 0.08667 0.01270
 0.09333 0.01576
 0.10000 0.01925
 0.10667 0.02319
 0.11333 0.02760
 0.12000 0.03249
 0.12667 0.03788
 0.13333 0.04378
 0.14000 0.05019
 0.14667 0.05713
 0.15333 0.06459
 0.16000 0.07258
 0.16667 0.08111
 0.17333 0.09016
 0.18000 0.09973
 0.18667 0.10982
 0.19333 0.12042
 0.20000 0.13151
 0.20667 0.14309
 0.21333 0.15513
 0.22000 0.16763
 0.22667 0.18056
 0.23333 0.19391
 0.24000 0.20765
 0.24667 0.22177
 0.25333 0.23623
 0.26000 0.25102
 0.26667 0.26611
 0.27333 0.28148
 0.28000 0.29709
 0.28667 0.31292
 0.29333 0.32895
 0.30000 0.34514
 0.30667 0.36146
 0.31333 0.37790
 0.32000 0.39442
 0.32667 0.41099
 0.33333 0.42759
 0.34000 0.44419
 0.34667 0.46077
 0.35333 0.47729
 0.36000 0.49373
 0.36667 0.51008
 0.37333 0.52630
 0.38000 0.54237
 0.38667 0.55828
 0.39333 0.57399
 0.40000 0.58950
 0.40667 0.60478
 0.41333 0.61982
 0.42000 0.63461
 0.42667 0.64912
 0.43333 0.66334
 0.44000 0.67726
 0.44667 0.69088
 0.45333 0.70417
 0.46000 0.71714
 0.46667 0.72977
 0.47333 0.74206
 0.48000 0.75400
 0.48667 0.76559
 0.49333 0.77683
 0.50000 0.78771
 0.50667 0.79824
 0.51333 0.80840
 0.52000 0.81822
 0.52667 0.82767
 0.53333 0.83678
 0.54000 0.84554
 0.54667 0.85395
 0.55333 0.86203
 0.56000 0.86977
 0.56667 0.87718
 0.57333 0.88427
 0.58000 0.89104
 0.58667 0.89751
 0.59333 0.90367
 0.60000 0.90955
 0.60667 0.91513
 0.61333 0.92044
 0.62000 0.92548
 0.62667 0.93026
 0.63333 0.93479
 0.64000 0.93908
 0.64667 0.94314
 0.65333 0.94697
 0.66000 0.95058
 0.66667 0.95399
 0.67333 0.95720
 0.68000 0.96022
 0.68667 0.96305
 0.69333 0.96572
 0.70000 0.96822
 0.70667 0.97056
 0.71333 0.97275
 0.72000 0.97480
 0.72667 0.97672
 0.73333 0.97851
 0.74000 0.98018
 0.74667 0.98173
 0.75333 0.98318
 0.76000 0.98452
 0.76667 0.98577
 0.77333 0.98693
 0.78000 0.98801
 0.78667 0.98900
 0.79333 0.98993
 0.80000 0.99078
 0.80667 0.99157
 0.81333 0.99229
 0.82000 0.99296
 0.82667 0.99358
 0.83333 0.99415
 0.84000 0.99467
 0.84667 0.99515
 0.85333 0.99559
 0.86000 0.99599
 0.86667 0.99636
 0.87333 0.99670
 0.88000 0.99701
 0.88667 0.99729
 0.89333 0.99755
 0.90000 0.99779
 0.90667 0.99800
 0.91333 0.99820
 0.92000 0.99837
 0.92667 0.99853
 0.93333 0.99868
 0.94000 0.99881
 0.94667 0.99893
 0.95333 0.99904
 0.96000 0.99914
 0.96667 0.99923
 0.97333 0.99931
 0.98000 0.99938
 0.98667 0.99945
 0.99333 0.99951
 1.00000 0.99956
/
{\small
\put
{\begin{tabular}{l}
normal\\distribution\\
obtained\\distribution
\end{tabular}} [l]
at 0.6 0.35
\putrule <0pt,1.5\baselineskip> from 0.5 0.35 to 0.6 0.35
\linethickness=\aaa\linethickness
\putrule <0pt,-0.5\baselineskip> from 0.5 0.35 to 0.6 0.35 }
\endpicture
}
\caption[]{Comparison of the angular momentum cumulative
distribution at a fixed mass with the normal distribution
(Eq.~(\ref{14})) for $M\gg M_{\rm b}$. {\bf a}~The region of the distribution tail
($\log_{10}M=1.8$) for $u=3/2$. The distribution is close to the
normal one. {\bf b}~cD-galaxies for $u=3/2$. The distribution
differs from the normal one (the significance level in the
Kolmogorov--Smirnov test is~$\sim10^{-9}$).}
\label{fig6}
\typeout{End drawing.}%
\end{figure}

The obtained momentum distribution at fixed mass in
the asymptotical region of large masses
is close to the normal distribution (Fig.~\ref{fig6}a). Thus, the
distribution tail may be represented as
\begin{equation}
f(M,\vec{S})\approx\Phi(M)
\left(\frac{2\pi}{3}\overline{S^2}\right)^{-3/2}
\exp\left(-\frac32\frac{S^2}{\overline{S^2}}\right),
\label{14}
\end{equation}
where $\Phi(M)$ is the mass function, $\overline{S^2(M)}$ is the
average square value of the momentum~$S$ for given mass~$M$.
cD-galaxies in the cases $u=3/2$ and $u=2$ make an exception
(Fig.~\ref{fig6}b). Kolmogorov--Smirnov test shows evidently that
their momentum distribution differs from the normal one.

\begin{figure}
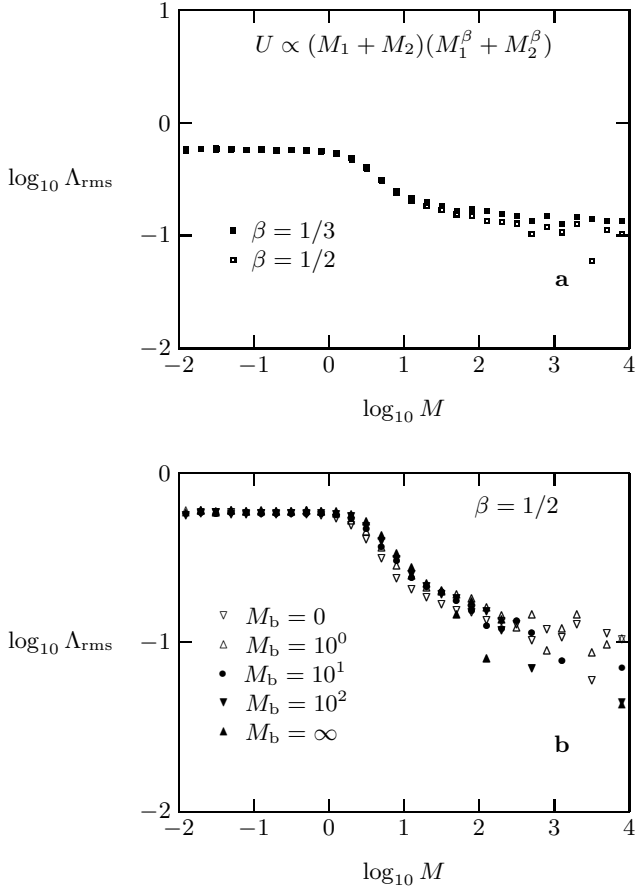

\typeout{Drawing figure...}%
\centerline{%
\beginpicture
\setcoordinatesystem units <\FigWidth,\FigHeight>
\setplotarea x from 0 to 1, y from 0 to 1
\axis bottom label {$\log_{10} M$} ticks in long numbered
  withvalues $-2$ $-1$ $0$ $1$ $2$ $3$ $4$ /
  quantity 7 /
\axis left label {$\log_{10} \Lambda_{\rm rms}$} ticks in long numbered
  withvalues $-2$ $-1$ $0$ $1$ /
  quantity 4 /
\axis top ticks in long
  quantity 7 /
\axis right ticks in long quantity 4 /
\put {\begin{tabular}{l}
 \raisebox{2pt}{\dimaplota}~~$\beta=1/3$\\
 \raisebox{2pt}{\dimaplotb}~~$\beta=1/2$
 \end{tabular}} at 0.25 0.3
\put {$U\propto(M_1+M_2)(M_1^\beta+M_2^\beta)$} at 0.5 0.9
\put {\bf a} at 0.85 0.2
\plotsymbolspacing50cm\setplotsymbol({\dimaplota})
\plot
0.01667  0.58854
0.05000  0.58803
0.08333  0.58696
0.11667  0.58718
0.15000  0.58608
0.18333  0.58741
0.21667  0.58600
0.25000  0.58697
0.28333  0.58544
0.31667  0.58182
0.35000  0.57452
0.38333  0.55767
0.41667  0.52885
0.45000  0.49650
0.48333  0.46575
0.51667  0.44519
0.55000  0.43158
0.58333  0.42178
0.61667  0.40685
0.65000  0.41286
0.68333  0.40595
0.71667  0.39903
0.75000  0.39184
0.78333  0.37647
0.81667  0.39136
0.85000  0.36858
0.88333  0.38883
0.91667  0.38228
0.95000  0.37562
0.98333  0.37600
/
\setplotsymbol({\dimaplotb})\plot
0.01667  0.58379
0.05000  0.58869
0.08333  0.59040
0.11667  0.58861
0.15000  0.58600
0.18333  0.58768
0.21667  0.58545
0.25000  0.58729
0.28333  0.58558
0.31667  0.58412
0.35000  0.57706
0.38333  0.56337
0.41667  0.53579
0.45000  0.49837
0.48333  0.45976
0.51667  0.43816
0.55000  0.42156
0.58333  0.40821
0.61667  0.39612
0.65000  0.39274
0.68333  0.37689
0.71667  0.37466
0.75000  0.36931
0.78333  0.33787
0.81667  0.35958
0.85000  0.34303
0.88333  0.36816
0.91667  0.25927
0.95000  0.35039
0.98333  0.33908
/
\endpicture
}
\figskip
\centerline{%
\beginpicture
\setcoordinatesystem units <\FigWidth,\FigHeight>
\setplotarea x from 0 to 1, y from 0 to 1
\axis bottom label {$\log_{10} M$} ticks in long numbered
  withvalues $-2$ $-1$ $0$ $1$ $2$ $3$ $4$ /
  quantity 7 /
\axis left label {$\log_{10} \Lambda_{\rm rms}$} ticks in long numbered
  withvalues $-2$ $-1$ $0$ /
  quantity 3 /
\axis top ticks in long
  quantity 7 /
\axis right ticks in long quantity 3 /
\put {\begin{tabular}{l}
 \raisebox{4pt}{\dimaplotf}~~$M_{\rm b}=0$\\
 \raisebox{0pt}{\dimaplotg}~~$M_{\rm b}=10^0$\\
 \raisebox{2pt}{\dimaplotc}~~$M_{\rm b}=10^1$\\
 \raisebox{4pt}{\dimaplotd}~~$M_{\rm b}=10^2$\\
 \raisebox{0pt}{\dimaplote}~~$M_{\rm b}=\infty$
 \end{tabular}} at 0.25 0.4
\put {$\beta=1/2$} at 0.75 0.9
\put {\bf b} at 0.85 0.2
\plotsymbolspacing50cm\setplotsymbol({\dimaplotf})
\plot
0.01667  0.87569
0.05000  0.88304
0.08333  0.88560
0.11667  0.88291
0.15000  0.87900
0.18333  0.88152
0.21667  0.87817
0.25000  0.88094
0.28333  0.87836
0.31667  0.87618
0.35000  0.86559
0.38333  0.84506
0.41667  0.80369
0.45000  0.74755
0.48333  0.68964
0.51667  0.65725
0.55000  0.63234
0.58333  0.61231
0.61667  0.59418
0.65000  0.58911
0.68333  0.56534
0.71667  0.56199
0.75000  0.55397
0.78333  0.50680
0.81667  0.53937
0.85000  0.51454
0.88333  0.55224
0.91667  0.38890
0.95000  0.52559
0.98333  0.50861
/
\setplotsymbol({\dimaplotg})
\plot
0.01667  0.88217
0.05000  0.88306
0.08333  0.87629
0.11667  0.88145
0.15000  0.88082
0.18333  0.88147
0.21667  0.88073
0.25000  0.88042
0.28333  0.88109
0.31667  0.87950
0.35000  0.87315
0.38333  0.85423
0.41667  0.82162
0.45000  0.77093
0.48333  0.71845
0.51667  0.69072
0.55000  0.65590
0.58333  0.64310
0.61667  0.63376
0.65000  0.62261
0.68333  0.59449
0.71667  0.57319
0.75000  0.53768
0.78333  0.57532
0.81667  0.46911
0.85000  0.53443
0.88333  0.57534
0.91667  0.46226
0.95000  0.48524
0.98333  0.50302
/
\setplotsymbol({\dimaplotc})
\plot
0.01667  0.87858
0.05000  0.88118
0.08333  0.88016
0.11667  0.88438
0.15000  0.88161
0.18333  0.88035
0.21667  0.87928
0.25000  0.87985
0.28333  0.88105
0.31667  0.87885
0.35000  0.87596
0.38333  0.86498
0.41667  0.83606
0.45000  0.78222
0.48333  0.74026
0.51667  0.68950
0.55000  0.66334
0.58333  0.64914
0.61667  0.62422
0.65000  0.59327
0.68333  0.55053
0.71667  0.54133
0.75000  0.56399
0.78333  0.52901
0.85000  0.44545
0.98333  0.42595
/
\setplotsymbol({\dimaplotd})
\plot
0.01667  0.87625
0.05000  0.88073
0.08333  0.88249
0.11667  0.87874
0.15000  0.88102
0.18333  0.88034
0.21667  0.88194
0.25000  0.88131
0.28333  0.88036
0.31667  0.87987
0.35000  0.87779
0.38333  0.87127
0.41667  0.84319
0.45000  0.79690
0.48333  0.74616
0.51667  0.70043
0.55000  0.66462
0.58333  0.64167
0.61667  0.62843
0.65000  0.59643
0.68333  0.59253
0.71667  0.53478
0.78333  0.42246
0.98333  0.32258
/
\setplotsymbol({\dimaplote})
\plot
0.01667  0.87551
0.05000  0.88199
0.08333  0.87906
0.11667  0.88231
0.15000  0.88051
0.18333  0.88054
0.21667  0.88049
0.25000  0.87976
0.28333  0.88096
0.31667  0.87983
0.35000  0.87977
0.38333  0.87107
0.41667  0.84951
0.45000  0.80744
0.48333  0.75510
0.51667  0.71475
0.55000  0.66713
0.58333  0.64548
0.61667  0.57432
0.65000  0.61482
0.68333  0.44644
0.71667  0.55947
0.98333  0.30987
/
\endpicture
}
\caption[]{Dependence of the root mean square dimensionless
momentum $\Lambda_{\rm rms}$ on mass.
{\bf a}~$M\gg M_{\rm b}$.
In the left-hand part of the
plot the main contribution is given by small galaxies which have
never merged, $\Lambda_{\rm rms}$~being determined by the initial
distribution for them. For large galaxies momenta are determined
by mergers and do not depend on the initial conditions,
$\Lambda_{\rm rms}\approx{\rm const}$.
{\bf b}~The same for different~$M_{\rm b}$
($N_{\rm f}=10^{-1}N$). $\Lambda_{\rm rms}$
decreases with~$M$.}
\label{fig3}
\typeout{End drawing.}%
\end{figure}

For $U\propto(M_1+M_2)\*(M_1^\beta+M_2^\beta)$ simulation shows
that, irrespectively of the initial momentum, the root mean square
value of the dimensionless momentum $\Lambda_{\rm rms}$ becomes
constant at large masses:
\begin{equation}
\Lambda_{\rm rms}\approx{\rm const}\sim0.1,
\label{15}
\end{equation}
that is
\begin{equation}
\overline{S^2(M)}\sim0.01\left[MR\sqrt{\frac{2GM}{R}}
\right]^2
\label{16}
\end{equation}
(Figs.~\ref{fig3}a and \ref{fig1}). The distribution at small
masses depends on $f_0(M,\vec{S})$.

For $U\propto(M_1+M_2)^2$ the dimensionless momentum decreases
with mass (Figs.~\ref{fig3}b and \ref{fig1prime}).

The fact that $\Lambda_{\rm rms}\approx{\rm const}$ for
$U\propto(M_1+M_2)\*(M_1^\beta+M_2^\beta)$ at large masses has a
simple qualitative explanation.  Consider the change of the mass
and momentum due to mergers. As the mass function decreases
rapidly (i.e., the number of small galaxies is very large) and
$u_1=0$, it is natural to suppose that the main contribution to
the change of the mass $M$ and momentum $S$ of a given massive
galaxy is associated with accretion of small galaxies $\sim M_0\ll
M$ (as, e.g., in Kontorovich et~al.\ \cite{jetplet}).
However, in this case the situation is different.
The rate of changing $M$ and $S$ due to mergers with
low mass galaxies ($<M$) can be expressed as
\begin{eqnarray}
\shifttoleft\dot M=\int_0^MU(M,M_1)M_1\Phi(M_1)\,\dd M_1
\nonumber\\
\shifttoleft
\qquad\propto M^{u_2}\int_0^MM_1^{u_1+1}\Phi(M_1)\,\dd M_1,
\label{17}\\
\shifttoleft\dot{\overline{S^2}}=\int_0^MU(M,M_1)(\overline{S^2(M_1)}+
   \overline{J^2(M,M_1)})\Phi(M_1)\,\dd M_1
\nonumber\\
\shifttoleft\qquad\propto
   M^{u_2+1+\beta}\int_0^M
   M_1^{u_1+2}\Phi(M_1)\,\dd M_1,
\label{18}
\end{eqnarray}
since $\overline{S^2(M_1)}\ll\overline{J^2(M,M_1)}$,
$\overline{J^2(M,M_1)}\sim M_1^2GMR$ for $M_1\ll M$. In our case
$u_1=0$, the slope of the mass function $\alpha\sim2$. Therefore,
the main contribution to integral~(\ref{18}) is given by large
galaxies ($\sim M$); contribution of large and small galaxies in
Eq.~(\ref{17}) are of the same order.

Next we consider successive mergers of galaxies with equal large
masses $M$ and momenta $S_{\rm rms}(M)$. The new galaxy has a mass
$M'=2M$ and momentum\footnote
{The orbital momentum
$\overline{J^2}=\overline{(MM/(M+M))^2v^2p_\infty^2}$, the
impact parameter
$\overline{p_\infty^2}=\frac12(R+R)^2v_{\rm g}^2/v^2$,
so, $\overline{J^2}=MGMR$.}
$\overline{(S')^2}=2\overline{S^2}+\overline{J^2}=2\overline{S^2}+MGMR$.
Thus, the new value of the
dimensionless momentum
$\overline{(\Lambda')^2}=2^{-2-\beta}\overline{\Lambda^2}+2^{-4-\beta}$.
As a result of successive mergers, $\Lambda_{\rm rms}$ tends to
the equilibrium value $\Lambda_{\rm
rms}=\sqrt{\displaystyle\frac{2^{-4-\beta}}{1-2^{-2-\beta}}}
\approx0.23\mbox{--}0.25$
($\beta=\frac13\mbox{--}\frac12$). So, the  result
$\overline{\Lambda^2}\approx{\rm const}$ is a consequence of the
fact that the main contribution to the change of mass and momentum
is given by mergers between comparable mass galaxies. The obtained
value for $\Lambda_{\rm rms}$ is somewhat different from
Eq.~(\ref{15}) due to simplifying assumptions made in the
derivation. Note, that for the isotropic momentum distribution
without allowance for the orbital momentum $\overline{S^2}\propto
M$, $\overline{\Lambda^2}\propto M^{-2-\beta}$; for the
anisotropic distribution (Kats \& Kontorovich
\cite{kkjetp}, \cite{kkpazh})
$\overline{S}\propto M$, $\overline{\Lambda}\propto
M^{-(1+\beta)/2}$.

In the case $U\propto(M_1+M_2)^2$
the main contribution to Eqs.~(\ref{17}), (\ref{18}) is given
by small masses $\sim M_0$ (due to large~$\alpha$) and
we can expect decreasing of
$\Lambda_{\rm rms}$ as~$M^{-1/2}$. This
can be demonstrated as follows. If integral $\int_0^\infty
M^{u_1+2}\Phi(M)\,\dd M$ converge at infinity then it is possible
to replace the upper limit of integrals (\ref{17}), (\ref{18}) by
infinity. Then
\begin{equation}
\dot M\propto M^{u_2},\qquad
\dot {\overline{S^2}}\propto M^{u_2+1+\beta}.
\label{19-20}
\end{equation}
Therefore,
\begin{equation}
\frac{\dd\overline{S^2}}{\dd M}\propto M^{1+\beta},\quad
\overline{S^2}\propto M^{2+\beta},\quad
\overline{\Lambda^2}\propto\frac{\overline{S^2}}{M^{3+\beta}}
 \propto M^{-1}.
\label{21-23}
\end{equation}

An analysis of observational data
(Kontorovich \& Khodjachikh \cite{konkhod}; Kontorovich et~al.\
\cite{kkgb})
confirms that $S_{\rm rms}\propto M^k$,
the coefficient $k$ being rather close to $(3+\beta)/2$, which is
in accordance with Eq.~(\ref{16}).

Note that allowance for dependence of the merger probability on
momenta may give an increase of~$\overline{\Lambda^2}$: numerical
experiments show that a merger is more probable if all momenta have
the same direction (Farouki \&
Shapiro \cite{fs82}). On the other hand, $\overline{\Lambda^2}$ is
sensitive to the exact form of the merger cross-section (in
particular, to the value of $\zeta$ in Eq.~(\ref{3})).

\section{Comparison of simulation results with
solution of the Smoluchowski equation}
\label{sec3}
Integrating the generalized Smoluchowski equation (\ref{1}) over
momenta, one can obtain the ordinary Smoluchowski
equation
which describe the evolution of the galaxy mass function:
\begin{eqnarray}
\shifttoleft{\partial \Phi(M,t)\over\partial t}
\nonumber\\
\shifttoleft\qquad=\int_0^M
U(M_1,M-M_1,t)\Phi(M_1,t)\Phi(M-M_1,t)\,\dd M_1
\nonumber\\
\shifttoleft\qquad-2\Phi(M,t)\int_0^\infty
U(M_1,M,t)\Phi(M_1,t)\,\dd M_1.
\label{24}
\end{eqnarray}
Solving this equation is another independent way
to find~$\Phi(M,t)$. In this section we compare the results
obtained by Monte Carlo simulation with the obtained earlier (see
Kontorovich et~al.\  \cite{physd} and Krivitsky
\cite{jphys})
results of direct numerical solving the
Smoluchowski equation with kernels
$U\propto(M_1+M_2)\*(M_1^\beta+M_2^\beta)$ and
$U\propto(M_1+M_2)^2$.

\begin{figure}
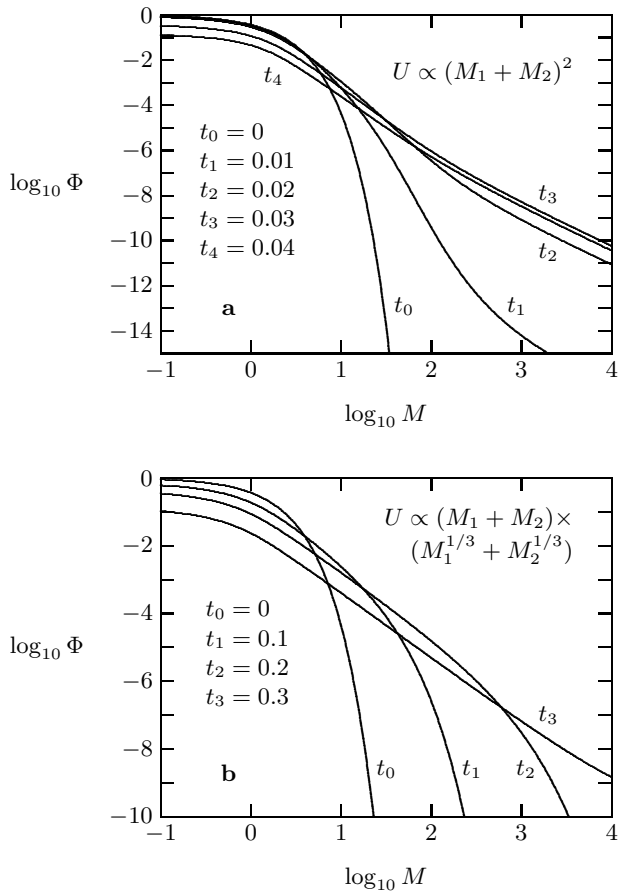

\typeout{Drawing figure...}%
\centerline{%
\beginpicture
\setcoordinatesystem units <\FigWidth,\FigHeight>
\setplotarea x from 0 to 1, y from 0 to 1
\axis bottom label {$\log_{10} M$} ticks in long numbered
  withvalues $-1$ $0$ $1$ $2$ $3$ $4$ /
  quantity 6 /
\axis left label {$\log_{10} \Phi$} ticks in long numbered
  withvalues {} $-14$ {} $-12$ {} $-10$
     {} $-8$ {} $-6$ {} $-4$ {} $-2$ {} $0$  /
  quantity 16 /
\axis top ticks in long
  quantity 6 /
\axis right ticks in long quantity 16 /
\put {$t_0$} at 0.54  0.14
\put {$t_1$} at 0.79  0.14
\put {$t_2$} at 0.86  0.3
\put {$t_3$} at 0.86  0.47
\put {$t_4$} at 0.25  0.82
\put {\bf a} at 0.15  0.13
\put
{\begin{tabular}{l}$t_0=0$\\$t_1=0.01$\\$t_2=0.02$\\$t_3=0.03$\\$t_4=0.04$\end{tabular}}
at 0.21 0.48
\put
{\begin{tabular}{r}$U\propto(M_1+M_2)^2$\end{tabular}}
[r] at 0.95 0.83
\plot
 0.00000  0.99710
 0.01859  0.99641
 0.03590  0.99562
 0.05194  0.99474
 0.06698  0.99374
 0.08124  0.99262
 0.09487  0.99137
 0.10799  0.98996
 0.12067  0.98838
 0.13299  0.98661
 0.14501  0.98463
 0.15677  0.98240
 0.16831  0.97990
 0.17966  0.97709
 0.19085  0.97394
 0.20189  0.97041
 0.21281  0.96645
 0.22362  0.96200
 0.23434  0.95701
 0.24497  0.95141
 0.25553  0.94513
 0.26603  0.93808
 0.27647  0.93017
 0.28686  0.92129
 0.29721  0.91134
 0.30752  0.90017
 0.31779  0.88763
 0.32803  0.87357
 0.33825  0.85779
 0.34844  0.84008
 0.35861  0.82021
 0.36876  0.79792
 0.37890  0.77291
 0.38902  0.74485
 0.39913  0.71336
 0.40922  0.67804
 0.41931  0.63840
 0.42938  0.59392
 0.43945  0.54402
 0.44951  0.48803
 0.45956  0.42521
 0.46961  0.35472
 0.47965  0.27563
 0.48969  0.18689
 0.49973  0.08732
 0.50756  0.00000
/
\plot
 0.00000  0.99568
 0.01859  0.99496
 0.03590  0.99413
 0.05194  0.99321
 0.06698  0.99216
 0.08124  0.99099
 0.09487  0.98968
 0.10799  0.98820
 0.12067  0.98655
 0.13299  0.98468
 0.14501  0.98259
 0.15677  0.98025
 0.16831  0.97762
 0.17966  0.97466
 0.19085  0.97135
 0.20189  0.96764
 0.21281  0.96349
 0.22362  0.95885
 0.23434  0.95368
 0.24497  0.94792
 0.25553  0.94154
 0.26603  0.93448
 0.27647  0.92673
 0.28686  0.91826
 0.29721  0.90908
 0.30752  0.89921
 0.31779  0.88868
 0.32803  0.87756
 0.33825  0.86591
 0.34844  0.85378
 0.35861  0.84118
 0.36876  0.82812
 0.37890  0.81459
 0.38902  0.80056
 0.39913  0.78601
 0.40922  0.77090
 0.41931  0.75521
 0.42938  0.73890
 0.43945  0.72196
 0.44951  0.70436
 0.45956  0.68607
 0.46961  0.66708
 0.47965  0.64739
 0.48969  0.62700
 0.49973  0.60592
 0.50976  0.58419
 0.51978  0.56186
 0.52981  0.53899
 0.53983  0.51568
 0.54985  0.49203
 0.55986  0.46817
 0.56988  0.44423
 0.57989  0.42037
 0.58990  0.39672
 0.59991  0.37342
 0.60992  0.35059
 0.61993  0.32833
 0.62994  0.30674
 0.63995  0.28589
 0.64995  0.26581
 0.65996  0.24654
 0.66996  0.22811
 0.67997  0.21050
 0.68997  0.19370
 0.69997  0.17771
 0.70998  0.16248
 0.71998  0.14799
 0.72998  0.13420
 0.73998  0.12106
 0.74999  0.10854
 0.75999  0.09659
 0.76999  0.08517
 0.77999  0.07424
 0.78999  0.06376
 0.79999  0.05369
 0.80999  0.04399
 0.81999  0.03463
 0.82999  0.02558
 0.84000  0.01682
 0.85000  0.00830
 0.86000  0.00002
 0.86002  0.00000
/
\plot
 0.00000  0.99298
 0.01859  0.99223
 0.03590  0.99137
 0.05194  0.99040
 0.06698  0.98931
 0.08124  0.98809
 0.09487  0.98672
 0.10799  0.98517
 0.12067  0.98344
 0.13299  0.98148
 0.14501  0.97929
 0.15677  0.97683
 0.16831  0.97407
 0.17966  0.97097
 0.19085  0.96750
 0.20189  0.96362
 0.21281  0.95929
 0.22362  0.95446
 0.23434  0.94911
 0.24497  0.94319
 0.25553  0.93670
 0.26603  0.92961
 0.27647  0.92195
 0.28686  0.91374
 0.29721  0.90505
 0.30752  0.89596
 0.31779  0.88654
 0.32803  0.87687
 0.33825  0.86700
 0.34844  0.85695
 0.35861  0.84675
 0.36876  0.83638
 0.37890  0.82586
 0.38902  0.81518
 0.39913  0.80434
 0.40922  0.79335
 0.41931  0.78220
 0.42938  0.77091
 0.43945  0.75947
 0.44951  0.74790
 0.45956  0.73620
 0.46961  0.72438
 0.47965  0.71246
 0.48969  0.70046
 0.49973  0.68839
 0.50976  0.67629
 0.51978  0.66417
 0.52981  0.65207
 0.53983  0.64001
 0.54985  0.62803
 0.55986  0.61615
 0.56988  0.60442
 0.57989  0.59285
 0.58990  0.58148
 0.59991  0.57033
 0.60992  0.55941
 0.61993  0.54874
 0.62994  0.53834
 0.63995  0.52819
 0.64995  0.51831
 0.65996  0.50870
 0.66996  0.49934
 0.67997  0.49022
 0.68997  0.48134
 0.69997  0.47268
 0.70998  0.46423
 0.71998  0.45598
 0.72998  0.44790
 0.73998  0.44000
 0.74999  0.43225
 0.75999  0.42464
 0.76999  0.41715
 0.77999  0.40979
 0.78999  0.40253
 0.79999  0.39537
 0.80999  0.38830
 0.81999  0.38131
 0.82999  0.37439
 0.84000  0.36753
 0.85000  0.36073
 0.86000  0.35399
 0.87000  0.34729
 0.88000  0.34063
 0.89000  0.33401
 0.90000  0.32743
 0.91000  0.32087
 0.92000  0.31435
 0.93000  0.30784
 0.94000  0.30136
 0.95000  0.29490
 0.96000  0.28846
 0.97000  0.28203
 0.98000  0.27562
 0.99000  0.26922
 1.00000  0.26283
/
\plot
 0.00000  0.96768
 0.01859  0.96691
 0.03590  0.96602
 0.05194  0.96503
 0.06698  0.96391
 0.08124  0.96264
 0.09487  0.96123
 0.10799  0.95963
 0.12067  0.95784
 0.13299  0.95582
 0.14501  0.95356
 0.15677  0.95101
 0.16831  0.94816
 0.17966  0.94495
 0.19085  0.94137
 0.20189  0.93736
 0.21281  0.93289
 0.22362  0.92793
 0.23434  0.92244
 0.24497  0.91640
 0.25553  0.90981
 0.26603  0.90268
 0.27647  0.89504
 0.28686  0.88695
 0.29721  0.87849
 0.30752  0.86974
 0.31779  0.86080
 0.32803  0.85172
 0.33825  0.84254
 0.34844  0.83328
 0.35861  0.82395
 0.36876  0.81455
 0.37890  0.80508
 0.38902  0.79556
 0.39913  0.78599
 0.40922  0.77638
 0.41931  0.76673
 0.42938  0.75705
 0.43945  0.74735
 0.44951  0.73764
 0.45956  0.72793
 0.46961  0.71823
 0.47965  0.70855
 0.48969  0.69890
 0.49973  0.68929
 0.50976  0.67974
 0.51978  0.67026
 0.52981  0.66085
 0.53983  0.65154
 0.54985  0.64233
 0.55986  0.63322
 0.56988  0.62423
 0.57989  0.61536
 0.58990  0.60662
 0.59991  0.59801
 0.60992  0.58954
 0.61993  0.58119
 0.62994  0.57297
 0.63995  0.56488
 0.64995  0.55691
 0.65996  0.54906
 0.66996  0.54133
 0.67997  0.53370
 0.68997  0.52617
 0.69997  0.51874
 0.70998  0.51140
 0.71998  0.50413
 0.72998  0.49695
 0.73998  0.48983
 0.74999  0.48278
 0.75999  0.47579
 0.76999  0.46886
 0.77999  0.46197
 0.78999  0.45513
 0.79999  0.44833
 0.80999  0.44158
 0.81999  0.43485
 0.82999  0.42816
 0.84000  0.42150
 0.85000  0.41487
 0.86000  0.40827
 0.87000  0.40169
 0.88000  0.39513
 0.89000  0.38859
 0.90000  0.38208
 0.91000  0.37558
 0.92000  0.36910
 0.93000  0.36264
 0.94000  0.35620
 0.95000  0.34977
 0.96000  0.34336
 0.97000  0.33697
 0.98000  0.33059
 0.99000  0.32422
 1.00000  0.31788
/
\plot
 0.00000  0.94066
 0.01859  0.93987
 0.03590  0.93898
 0.05194  0.93797
 0.06698  0.93683
 0.08124  0.93556
 0.09487  0.93412
 0.10799  0.93251
 0.12067  0.93069
 0.13299  0.92865
 0.14501  0.92636
 0.15677  0.92378
 0.16831  0.92089
 0.17966  0.91765
 0.19085  0.91402
 0.20189  0.90996
 0.21281  0.90544
 0.22362  0.90043
 0.23434  0.89488
 0.24497  0.88880
 0.25553  0.88217
 0.26603  0.87501
 0.27647  0.86737
 0.28686  0.85930
 0.29721  0.85090
 0.30752  0.84225
 0.31779  0.83343
 0.32803  0.82450
 0.33825  0.81550
 0.34844  0.80643
 0.35861  0.79731
 0.36876  0.78814
 0.37890  0.77893
 0.38902  0.76969
 0.39913  0.76042
 0.40922  0.75113
 0.41931  0.74182
 0.42938  0.73251
 0.43945  0.72320
 0.44951  0.71390
 0.45956  0.70462
 0.46961  0.69537
 0.47965  0.68616
 0.48969  0.67699
 0.49973  0.66788
 0.50976  0.65884
 0.51978  0.64987
 0.52981  0.64098
 0.53983  0.63217
 0.54985  0.62346
 0.55986  0.61485
 0.56988  0.60635
 0.57989  0.59794
 0.58990  0.58965
 0.59991  0.58146
 0.60992  0.57338
 0.61993  0.56540
 0.62994  0.55752
 0.63995  0.54974
 0.64995  0.54206
 0.65996  0.53446
 0.66996  0.52695
 0.67997  0.51952
 0.68997  0.51217
 0.69997  0.50489
 0.70998  0.49767
 0.71998  0.49052
 0.72998  0.48342
 0.73998  0.47638
 0.74999  0.46939
 0.75999  0.46244
 0.76999  0.45553
 0.77999  0.44866
 0.78999  0.44183
 0.79999  0.43503
 0.80999  0.42827
 0.81999  0.42154
 0.82999  0.41483
 0.84000  0.40815
 0.85000  0.40149
 0.86000  0.39486
 0.87000  0.38825
 0.88000  0.38167
 0.89000  0.37510
 0.90000  0.36856
 0.91000  0.36203
 0.92000  0.35553
 0.93000  0.34905
 0.94000  0.34258
 0.95000  0.33614
 0.96000  0.32971
 0.97000  0.32331
 0.98000  0.31692
 0.99000  0.31055
 1.00000  0.30421
/
\endpicture
}
\figskip
\centerline{%
\beginpicture
\setcoordinatesystem units <\FigWidth,\FigHeight>
\setplotarea x from 0 to 1, y from 0 to 1
\axis bottom label {$\log_{10} M$} ticks in long numbered
  withvalues $-1$ $0$ $1$ $2$ $3$ $4$ /
  quantity 6 /
\axis left label {$\log_{10} \Phi$} ticks in long numbered
  withvalues $-10$
     {} $-8$ {} $-6$ {} $-4$ {} $-2$ {} $0$  /
  quantity 11 /
\axis top ticks in long
  quantity 6 /
\axis right ticks in long quantity 11 /
\put {$t_0$} at 0.50  0.14
\put {$t_1$} at 0.69  0.14
\put {$t_2$} at 0.81  0.14
\put {$t_3$} at 0.86  0.3
\put {\bf b} at 0.15  0.13
\put
{\begin{tabular}{l}$t_0=0$\\$t_1=0.1$\\$t_2=0.2$\\$t_3=0.3$\end{tabular}}
at 0.21 0.48
\put
{\begin{tabular}{r}$U\propto(M_1+M_2)\times$\\$(M_1^{1/3}+M_2^{1/3})$\end{tabular}}
[r] at 0.95 0.83
\plot
 0.00000  0.99566
 0.00934  0.99516
 0.01859  0.99462
 0.02743  0.99404
 0.03590  0.99343
 0.04406  0.99279
 0.05194  0.99210
 0.05957  0.99138
 0.06698  0.99061
 0.07420  0.98980
 0.08124  0.98893
 0.08813  0.98802
 0.09487  0.98705
 0.10149  0.98603
 0.10799  0.98494
 0.11438  0.98379
 0.12067  0.98258
 0.12687  0.98129
 0.13299  0.97992
 0.13904  0.97847
 0.14501  0.97694
 0.15092  0.97532
 0.15677  0.97360
 0.16257  0.97178
 0.16831  0.96985
 0.17401  0.96780
 0.17966  0.96564
 0.18527  0.96334
 0.19085  0.96091
 0.19639  0.95834
 0.20189  0.95561
 0.20737  0.95273
 0.21281  0.94967
 0.21823  0.94643
 0.22362  0.94300
 0.22899  0.93936
 0.23434  0.93551
 0.23967  0.93143
 0.24497  0.92711
 0.25026  0.92254
 0.25553  0.91769
 0.26079  0.91256
 0.26603  0.90712
 0.27126  0.90136
 0.27647  0.89525
 0.28167  0.88879
 0.28686  0.88194
 0.29204  0.87469
 0.29721  0.86701
 0.30237  0.85887
 0.30752  0.85025
 0.31266  0.84112
 0.31779  0.83145
 0.32292  0.82120
 0.32803  0.81035
 0.33314  0.79886
 0.33825  0.78668
 0.34335  0.77378
 0.34844  0.76012
 0.35353  0.74565
 0.35861  0.73032
 0.36369  0.71408
 0.36876  0.69688
 0.37383  0.67867
 0.37890  0.65937
 0.38396  0.63893
 0.38902  0.61728
 0.39408  0.59434
 0.39913  0.57005
 0.40418  0.54431
 0.40922  0.51705
 0.41427  0.48818
 0.41931  0.45760
 0.42435  0.42520
 0.42938  0.39088
 0.43442  0.35453
 0.43945  0.31603
 0.44448  0.27524
 0.44951  0.23204
 0.45454  0.18628
 0.45956  0.13781
 0.46459  0.08646
 0.46961  0.03207
 0.47241  0.00000
/
\plot
 0.00000  0.97852
 0.00934  0.97776
 0.01859  0.97694
 0.02743  0.97608
 0.03590  0.97517
 0.04406  0.97423
 0.05194  0.97323
 0.05957  0.97218
 0.06698  0.97109
 0.07420  0.96993
 0.08124  0.96872
 0.08813  0.96744
 0.09487  0.96610
 0.10149  0.96469
 0.10799  0.96320
 0.11438  0.96164
 0.12067  0.96000
 0.12687  0.95828
 0.13299  0.95646
 0.13904  0.95456
 0.14501  0.95256
 0.15092  0.95046
 0.15677  0.94826
 0.16257  0.94595
 0.16831  0.94353
 0.17401  0.94100
 0.17966  0.93835
 0.18527  0.93559
 0.19085  0.93270
 0.19639  0.92969
 0.20189  0.92655
 0.20737  0.92328
 0.21281  0.91989
 0.21823  0.91637
 0.22362  0.91273
 0.22899  0.90896
 0.23434  0.90507
 0.23967  0.90107
 0.24497  0.89695
 0.25026  0.89272
 0.25553  0.88838
 0.26079  0.88395
 0.26603  0.87942
 0.27126  0.87481
 0.27647  0.87011
 0.28167  0.86535
 0.28686  0.86051
 0.29204  0.85560
 0.29721  0.85064
 0.30237  0.84562
 0.30752  0.84054
 0.31266  0.83541
 0.31779  0.83023
 0.32292  0.82500
 0.32803  0.81972
 0.33314  0.81439
 0.33825  0.80901
 0.34335  0.80357
 0.34844  0.79808
 0.35353  0.79253
 0.35861  0.78693
 0.36369  0.78127
 0.36876  0.77554
 0.37383  0.76975
 0.37890  0.76389
 0.38396  0.75796
 0.38902  0.75196
 0.39408  0.74588
 0.39913  0.73972
 0.40418  0.73347
 0.40922  0.72714
 0.41427  0.72071
 0.41931  0.71419
 0.42435  0.70756
 0.42938  0.70082
 0.43442  0.69397
 0.43945  0.68700
 0.44448  0.67991
 0.44951  0.67268
 0.45454  0.66532
 0.45956  0.65781
 0.46459  0.65015
 0.46961  0.64232
 0.47463  0.63433
 0.47965  0.62616
 0.48467  0.61780
 0.48969  0.60924
 0.49471  0.60047
 0.49973  0.59149
 0.50474  0.58227
 0.50976  0.57281
 0.51477  0.56310
 0.51978  0.55311
 0.52479  0.54284
 0.52981  0.53228
 0.53482  0.52140
 0.53983  0.51019
 0.54484  0.49863
 0.54985  0.48670
 0.55485  0.47439
 0.55986  0.46167
 0.56487  0.44853
 0.56988  0.43493
 0.57488  0.42087
 0.57989  0.40630
 0.58490  0.39121
 0.58990  0.37557
 0.59491  0.35934
 0.59991  0.34251
 0.60492  0.32503
 0.60992  0.30688
 0.61493  0.28802
 0.61993  0.26841
 0.62494  0.24802
 0.62994  0.22680
 0.63494  0.20471
 0.63995  0.18171
 0.64495  0.15775
 0.64995  0.13279
 0.65495  0.10677
 0.65996  0.07964
 0.66496  0.05135
 0.66996  0.02184
 0.67351  0.00000
/
\plot
 0.00000  0.95397
 0.00934  0.95299
 0.01859  0.95194
 0.02743  0.95085
 0.03590  0.94971
 0.04406  0.94852
 0.05194  0.94728
 0.05957  0.94598
 0.06698  0.94463
 0.07420  0.94321
 0.08124  0.94172
 0.08813  0.94017
 0.09487  0.93855
 0.10149  0.93684
 0.10799  0.93506
 0.11438  0.93320
 0.12067  0.93124
 0.12687  0.92920
 0.13299  0.92706
 0.13904  0.92482
 0.14501  0.92249
 0.15092  0.92004
 0.15677  0.91750
 0.16257  0.91484
 0.16831  0.91207
 0.17401  0.90918
 0.17966  0.90618
 0.18527  0.90307
 0.19085  0.89984
 0.19639  0.89649
 0.20189  0.89303
 0.20737  0.88946
 0.21281  0.88578
 0.21823  0.88200
 0.22362  0.87811
 0.22899  0.87413
 0.23434  0.87006
 0.23967  0.86590
 0.24497  0.86167
 0.25026  0.85737
 0.25553  0.85301
 0.26079  0.84860
 0.26603  0.84413
 0.27126  0.83962
 0.27647  0.83508
 0.28167  0.83051
 0.28686  0.82591
 0.29204  0.82129
 0.29721  0.81665
 0.30237  0.81199
 0.30752  0.80732
 0.31266  0.80263
 0.31779  0.79793
 0.32292  0.79323
 0.32803  0.78851
 0.33314  0.78378
 0.33825  0.77904
 0.34335  0.77430
 0.34844  0.76955
 0.35353  0.76479
 0.35861  0.76002
 0.36369  0.75525
 0.36876  0.75046
 0.37383  0.74568
 0.37890  0.74088
 0.38396  0.73608
 0.38902  0.73127
 0.39408  0.72645
 0.39913  0.72162
 0.40418  0.71679
 0.40922  0.71195
 0.41427  0.70711
 0.41931  0.70225
 0.42435  0.69739
 0.42938  0.69252
 0.43442  0.68764
 0.43945  0.68276
 0.44448  0.67786
 0.44951  0.67296
 0.45454  0.66805
 0.45956  0.66312
 0.46459  0.65819
 0.46961  0.65325
 0.47463  0.64829
 0.47965  0.64333
 0.48467  0.63835
 0.48969  0.63336
 0.49471  0.62836
 0.49973  0.62334
 0.50474  0.61831
 0.50976  0.61327
 0.51477  0.60821
 0.51978  0.60314
 0.52479  0.59805
 0.52981  0.59294
 0.53482  0.58782
 0.53983  0.58267
 0.54484  0.57751
 0.54985  0.57233
 0.55485  0.56713
 0.55986  0.56190
 0.56487  0.55665
 0.56988  0.55138
 0.57488  0.54608
 0.57989  0.54076
 0.58490  0.53541
 0.58990  0.53003
 0.59491  0.52462
 0.59991  0.51918
 0.60492  0.51371
 0.60992  0.50820
 0.61493  0.50265
 0.61993  0.49707
 0.62494  0.49145
 0.62994  0.48579
 0.63494  0.48009
 0.63995  0.47434
 0.64495  0.46854
 0.64995  0.46270
 0.65495  0.45680
 0.65996  0.45085
 0.66496  0.44484
 0.66996  0.43878
 0.67496  0.43265
 0.67997  0.42646
 0.68497  0.42021
 0.68997  0.41388
 0.69497  0.40748
 0.69997  0.40100
 0.70497  0.39444
 0.70998  0.38780
 0.71498  0.38107
 0.71998  0.37425
 0.72498  0.36733
 0.72998  0.36032
 0.73498  0.35319
 0.73998  0.34596
 0.74498  0.33862
 0.74999  0.33115
 0.75499  0.32356
 0.75999  0.31584
 0.76499  0.30798
 0.76999  0.29999
 0.77499  0.29184
 0.77999  0.28353
 0.78499  0.27507
 0.78999  0.26643
 0.79499  0.25762
 0.79999  0.24863
 0.80499  0.23944
 0.80999  0.23005
 0.81499  0.22046
 0.81999  0.21064
 0.82499  0.20060
 0.82999  0.19032
 0.83499  0.17980
 0.84000  0.16902
 0.84500  0.15798
 0.85000  0.14666
 0.85500  0.13505
 0.86000  0.12315
 0.86500  0.11094
 0.87000  0.09840
 0.87500  0.08553
 0.88000  0.07232
 0.88500  0.05876
 0.89000  0.04483
 0.89500  0.03051
 0.90000  0.01581
 0.90500  0.00071
 0.90523  0.00000
/
\plot
 0.00000  0.90262
 0.00934  0.90149
 0.01859  0.90028
 0.02743  0.89902
 0.03590  0.89772
 0.04406  0.89636
 0.05194  0.89496
 0.05957  0.89349
 0.06698  0.89197
 0.07420  0.89038
 0.08124  0.88872
 0.08813  0.88699
 0.09487  0.88518
 0.10149  0.88330
 0.10799  0.88133
 0.11438  0.87927
 0.12067  0.87712
 0.12687  0.87488
 0.13299  0.87254
 0.13904  0.87010
 0.14501  0.86756
 0.15092  0.86491
 0.15677  0.86215
 0.16257  0.85928
 0.16831  0.85629
 0.17401  0.85320
 0.17966  0.84998
 0.18527  0.84666
 0.19085  0.84321
 0.19639  0.83966
 0.20189  0.83599
 0.20737  0.83222
 0.21281  0.82834
 0.21823  0.82437
 0.22362  0.82030
 0.22899  0.81614
 0.23434  0.81191
 0.23967  0.80760
 0.24497  0.80322
 0.25026  0.79878
 0.25553  0.79429
 0.26079  0.78976
 0.26603  0.78518
 0.27126  0.78057
 0.27647  0.77594
 0.28167  0.77128
 0.28686  0.76660
 0.29204  0.76190
 0.29721  0.75719
 0.30237  0.75247
 0.30752  0.74774
 0.31266  0.74300
 0.31779  0.73825
 0.32292  0.73349
 0.32803  0.72873
 0.33314  0.72396
 0.33825  0.71919
 0.34335  0.71441
 0.34844  0.70963
 0.35353  0.70485
 0.35861  0.70006
 0.36369  0.69527
 0.36876  0.69048
 0.37383  0.68568
 0.37890  0.68088
 0.38396  0.67608
 0.38902  0.67128
 0.39408  0.66648
 0.39913  0.66167
 0.40418  0.65686
 0.40922  0.65206
 0.41427  0.64725
 0.41931  0.64244
 0.42435  0.63763
 0.42938  0.63282
 0.43442  0.62800
 0.43945  0.62319
 0.44448  0.61838
 0.44951  0.61357
 0.45454  0.60875
 0.45956  0.60394
 0.46459  0.59913
 0.46961  0.59431
 0.47463  0.58950
 0.47965  0.58469
 0.48467  0.57988
 0.48969  0.57507
 0.49471  0.57026
 0.49973  0.56545
 0.50474  0.56064
 0.50976  0.55583
 0.51477  0.55102
 0.51978  0.54621
 0.52479  0.54141
 0.52981  0.53660
 0.53482  0.53180
 0.53983  0.52700
 0.54484  0.52220
 0.54985  0.51740
 0.55485  0.51260
 0.55986  0.50781
 0.56487  0.50301
 0.56988  0.49822
 0.57488  0.49343
 0.57989  0.48864
 0.58490  0.48385
 0.58990  0.47907
 0.59491  0.47429
 0.59991  0.46951
 0.60492  0.46473
 0.60992  0.45996
 0.61493  0.45519
 0.61993  0.45042
 0.62494  0.44566
 0.62994  0.44089
 0.63494  0.43613
 0.63995  0.43138
 0.64495  0.42663
 0.64995  0.42188
 0.65495  0.41713
 0.65996  0.41239
 0.66496  0.40766
 0.66996  0.40293
 0.67496  0.39820
 0.67997  0.39347
 0.68497  0.38875
 0.68997  0.38404
 0.69497  0.37933
 0.69997  0.37463
 0.70497  0.36993
 0.70998  0.36524
 0.71498  0.36055
 0.71998  0.35587
 0.72498  0.35120
 0.72998  0.34653
 0.73498  0.34187
 0.73998  0.33721
 0.74498  0.33256
 0.74999  0.32792
 0.75499  0.32329
 0.75999  0.31867
 0.76499  0.31405
 0.76999  0.30945
 0.77499  0.30485
 0.77999  0.30026
 0.78499  0.29568
 0.78999  0.29111
 0.79499  0.28655
 0.79999  0.28200
 0.80499  0.27747
 0.80999  0.27294
 0.81499  0.26843
 0.81999  0.26392
 0.82499  0.25944
 0.82999  0.25496
 0.83499  0.25050
 0.84000  0.24605
 0.84500  0.24162
 0.85000  0.23721
 0.85500  0.23281
 0.86000  0.22842
 0.86500  0.22406
 0.87000  0.21971
 0.87500  0.21539
 0.88000  0.21108
 0.88500  0.20679
 0.89000  0.20253
 0.89500  0.19829
 0.90000  0.19407
 0.90500  0.18987
 0.91000  0.18571
 0.91500  0.18156
 0.92000  0.17745
 0.92500  0.17337
 0.93000  0.16932
 0.93500  0.16529
 0.94000  0.16131
 0.94500  0.15736
 0.95000  0.15344
 0.95500  0.14956
 0.96000  0.14573
 0.96500  0.14194
 0.97000  0.13819
 0.97500  0.13448
 0.98000  0.13083
 0.98500  0.12723
 0.99000  0.12368
 0.99500  0.12019
 1.00000  0.11676
/
\endpicture
}
\caption[]{{\bf a}~Numerical solution of the Smoluchowski equation
in the small mass region
($M_{\rm max}=10^4$,
$M$~is measured in units~$M_0$, time in units $1/(c_uN_0M_0^u)$,
$\Phi$~in units~$N_0/M_0$).
{\bf b}~Solution in the large mass region, for comparison (the
same case as in Cavaliere et~al.\ (\cite{ccm92}), but in a wider
mass range and somewhat more precise).}
\label{fig4}
\typeout{End drawing.}%
\end{figure}

\begin{figure}
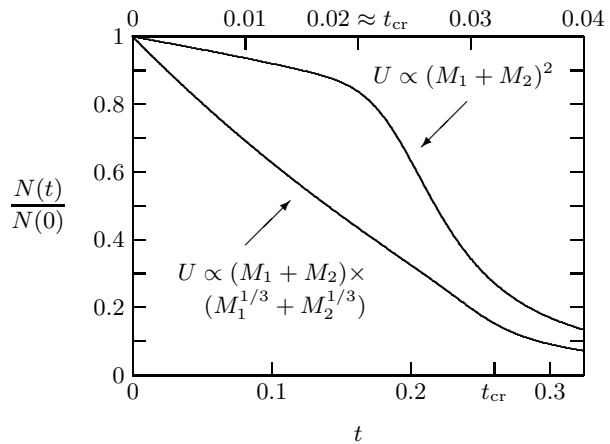

\typeout{Drawing figure...}%
\setlength{\unitlength}{1\FigWidth}
\centerline{%
\beginpicture
\setcoordinatesystem units <\FigWidth,\FigHeight>
\setplotarea x from 0 to 1, y from 0 to 1
\axis bottom label {$t$} ticks in long numbered
  withvalues $0$ $0.1$ $0.2$ $t_{\rm cr}$ $0.3$ /
  at 0 0.30864197 0.61728395 0.80246914 0.92592593 / /
\axis left label {$\displaystyle\frac{N(t)}{N(0)}$} ticks in long numbered
  withvalues $0$
     {} $0.2$ {} $0.4$ {} $0.6$ {} $0.8$ {} $1$  /
  quantity 11 /
\axis top ticks in long numbered withvalues
  $0$ $0.01$ {$0.02\approx t_{\rm cr}$} $0.03$ $0.04$ /
  quantity 5 /
\axis right ticks in long quantity 11 /
\put{\vector(-1,-1){0.1}} at 0.72 0.74
\put{\vector(1,1){0.1}} at 0.3 0.45
\put
{\begin{tabular}{r}$U\propto(M_1+M_2)^2$\end{tabular}}
[r] at 0.97 0.88
\put
{\begin{tabular}{r}$U\propto(M_1+M_2)\times$\\$(M_1^{1/3}+M_2^{1/3})$\end{tabular}}
[l] at 0.1 0.25
\plot
  0.00000  1.00000
  0.01019  0.98628
  0.02037  0.97269
  0.03056  0.95920
  0.04074  0.94582
  0.05093  0.93253
  0.06111  0.91935
  0.07130  0.90627
  0.08148  0.89328
  0.09167  0.88040
  0.10185  0.86761
  0.11204  0.85492
  0.12222  0.84232
  0.13241  0.82982
  0.14259  0.81742
  0.15278  0.80511
  0.16296  0.79289
  0.17315  0.78076
  0.18333  0.76873
  0.19352  0.75679
  0.20370  0.74494
  0.21389  0.73317
  0.22407  0.72150
  0.23426  0.70991
  0.24444  0.69842
  0.25463  0.68700
  0.26481  0.67567
  0.27500  0.66443
  0.28519  0.65327
  0.29537  0.64219
  0.30556  0.63119
  0.31574  0.62027
  0.32593  0.60943
  0.33611  0.59867
  0.34630  0.58798
  0.35648  0.57737
  0.36667  0.56683
  0.37685  0.55637
  0.38704  0.54598
  0.39722  0.53565
  0.40741  0.52540
  0.41759  0.51521
  0.42778  0.50508
  0.43796  0.49502
  0.44815  0.48502
  0.45833  0.47508
  0.46852  0.46520
  0.47870  0.45537
  0.48889  0.44559
  0.49907  0.43586
  0.50926  0.42618
  0.51944  0.41654
  0.52963  0.40694
  0.53981  0.39738
  0.55000  0.38785
  0.56019  0.37835
  0.57037  0.36887
  0.58056  0.35941
  0.59074  0.34996
  0.60093  0.34052
  0.61111  0.33107
  0.62130  0.32162
  0.63148  0.31214
  0.64167  0.30264
  0.65185  0.29309
  0.66204  0.28348
  0.67222  0.27380
  0.68241  0.26404
  0.69259  0.25418
  0.70278  0.24423
  0.71296  0.23421
  0.72315  0.22415
  0.73333  0.21411
  0.74352  0.20418
  0.75370  0.19444
  0.76389  0.18498
  0.77407  0.17588
  0.78426  0.16719
  0.79444  0.15897
  0.80463  0.15123
  0.81481  0.14398
  0.82500  0.13721
  0.83519  0.13092
  0.84537  0.12506
  0.85556  0.11962
  0.86574  0.11457
  0.87593  0.10988
  0.88611  0.10552
  0.89630  0.10146
  0.90648  0.09768
  0.91667  0.09416
  0.92685  0.09086
  0.93704  0.08778
  0.94722  0.08489
  0.95741  0.08218
  0.96759  0.07964
  0.97778  0.07724
  0.98796  0.07498
  1.00000  0.07248
/
\plot
  0.00000  1.00000
  0.01250  0.99690
  0.02500  0.99388
  0.03750  0.99085
  0.05000  0.98780
  0.06250  0.98474
  0.07500  0.98166
  0.08750  0.97856
  0.10000  0.97545
  0.11250  0.97232
  0.12500  0.96917
  0.13750  0.96601
  0.15000  0.96282
  0.16250  0.95961
  0.17500  0.95638
  0.18750  0.95313
  0.20000  0.94986
  0.21250  0.94656
  0.22500  0.94324
  0.23750  0.93989
  0.25000  0.93651
  0.26250  0.93310
  0.27500  0.92967
  0.28750  0.92620
  0.30000  0.92270
  0.31250  0.91916
  0.32500  0.91557
  0.33750  0.91194
  0.35000  0.90824
  0.36250  0.90447
  0.37500  0.90059
  0.38750  0.89658
  0.40000  0.89239
  0.41250  0.88794
  0.42500  0.88314
  0.43750  0.87787
  0.45000  0.87199
  0.46250  0.86530
  0.47500  0.85757
  0.48750  0.84857
  0.50000  0.83800
  0.51250  0.82560
  0.52500  0.81111
  0.53750  0.79430
  0.55000  0.77503
  0.56250  0.75328
  0.57500  0.72913
  0.58750  0.70279
  0.60000  0.67461
  0.61250  0.64501
  0.62500  0.61449
  0.63750  0.58355
  0.65000  0.55269
  0.66250  0.52232
  0.67500  0.49282
  0.68750  0.46446
  0.70000  0.43745
  0.71250  0.41190
  0.72500  0.38789
  0.73750  0.36542
  0.75000  0.34448
  0.76250  0.32501
  0.77500  0.30694
  0.78750  0.29020
  0.80000  0.27470
  0.81250  0.26034
  0.82500  0.24705
  0.83750  0.23474
  0.85000  0.22332
  0.86250  0.21273
  0.87500  0.20290
  0.88750  0.19376
  0.90000  0.18525
  0.91250  0.17733
  0.92500  0.16994
  0.93750  0.16303
  0.95000  0.15657
  0.96250  0.15052
  0.97500  0.14485
  0.98750  0.13953
  1.00000  0.13452
/
\endpicture
}
\caption[]{Time dependence of the number of galaxies, obtained
from the solution of the Smoluchowski equation
(upper time scale for $u=2$, lower time scale for $u=4/3$).
Though the total number of galaxies decreases, at any $t$ the
number of low-mass galaxies which have never merged is much less
than the number of massive tail galaxies.}
\label{fig5}
\typeout{End drawing.}%
\end{figure}

For numerical solution of Eq.~(\ref{24}) and analysis of the
obtained results  we used methods described in Krivitsky
\cite{jphys}; Kontorovich et~al.\ \cite{physd}. Figure~\ref{fig4}b
shows a plot of the obtained mass function for
$U=c_u(M_1+M_2)\*(M_1^\beta+M_2^\beta)$.  One can see that an
intermediate asymptotics, close to a power law~$M^{-\alpha}$, is
formed in the region $M_0\ll M\ll M_{\rm max}$. The exponent
$\alpha\approx1.9$ for $u=4/3$ and $\alpha\approx2.1$ for $u=3/2$
(in the latter case $\alpha$ is defined worse, see below). The
value $1.9$ differs from $\alpha\approx2.15$ given by Cavaliere
et~al.\ (\cite{ccm92}) but, in our opinion, agrees rather well
with the plot in their work. The value of $t_{\rm cr}$ for the two
cases are\footnote{Since a characteristic time
associated
with mergers is of the order of $(c_uN_0M_0^u)^{-1}$, such a value
of $t_{\rm cr}$ means that the phase transition is comparatively
fast. Due to considerable contribution of
a comparatively small number of appearing massive galaxies,
the time $t_{\rm cr}$ corresponding to the ``phase
transition'' is much less than a characteristic time $(\sigma
vn)^{-1}$, where $\sigma$ is the merger cross-section for typical
galaxies, $n$ is the concentration of such galaxies, $v$ is the
average velocity.}, respectively, $0.26$ and
$0.1$ (in the same units as in Fig.~\ref{fig4})
for the initial distribution
$\Phi_0(M)=(N_0/M_0)\ee^{-M/M_0}$.
The time dependence of the number of galaxies, obtained
from the solution of the Smoluchowski equation, is shown in
Fig.~\ref{fig5}. The moment displayed in Fig.~\ref{fig1}b ($N_{\rm
f}=0.1N_0$) approximately corresponds to $t\approx0.3$ (whereas
$t_{\rm cr}\approx0.26$).

In the case $U=c_2(M_1+M_2)^2$ (Fig.~\ref{fig4}a) the intermediate
asymptotics is not as close to a power law as for
$U=c_u(M_1+M_2)\*(M_1^\beta+M_2^\beta)$. An effective slope $\alpha$
in this case is~2--3. The value of $t_{\rm cr}$ is
$\sim0.02$, that is the phase transition is very
fast. However, in this case the time dependence of the
distribution moments is non-power and $t_{\rm cr}$ cannot be
determined accurately (Kontorovich et~al.\ \cite{physd}; Krivitsky
\cite{jphys}).

In numerical solving Eq.~(\ref{24}), a finite limit mass $M_{\rm
max}$ was introduced: the integral from 0 to $\infty$ in the
right-hand part was replaced by the integral from 0 to~$M_{\rm
max}$. Physically, such a substitution corresponds to a sink at
large masses. As it was shown by Krivitsky \cite{jphys},
Kontorovich et~al.\ \cite{physd}, consequences of this replacement
are different for kernels with $u_2\le1$ and $u_2>1$. In the case
of kernel~(\ref{7}) which belongs to the class $u_2>1$, existence
of $M_{\rm max}$ and its value essentially influence the solution,
in particular, the number of galaxies $N=\int_0^\infty\Phi\,\dd M$
and distribution moments ${\cal M}^{(p)}=\int_0^\infty\Phi
M^p\,\dd M$ as functions of time, the value of $t_{\rm cr}$ etc.
Moreover, van~Dongen (\cite{nosol}) showed that the limit $M_{\rm
max}\to\infty$ does not exist at all for $u_2>u_1+1$ (this is the
case for Eq.~(\ref{7})). The influence of $M_{\rm max}$ increases
as $u_2$ becomes farther from~1: for
$U\propto(M_1+M_2)\*(M_1^\beta+M_2^\beta)$, especially if
$\beta=1/3$, this influence is moderate. The farther $u_2$ moves
from~1, the greater the difference between the intermediate
asymptotics and the power law becomes and the worse defined
$t_{\rm cr}$ and $\alpha$ are; this is the case for
$U\propto(M_1+M_2)^2$.

As known (see  van~Dongen \& Ernst \cite{evd}; Voloshchuk
\cite{vol}; Krivitsky \cite{jphys}; Kontorovich et~al.\
\cite{physd}), in many cases the solution for $u_2\le1$ is
self-similar:  $\Phi(M)\approx\mu^{-\tau}(t)\varphi(M/\mu(t))$ for
$M\gg M_0$, where $\varphi$ is a time-independent function,
$\mu(t)$ describes a ``front'' moving to greater masses. The
numerical solution shows that $\Phi(M,t)$ for
$U\propto(M_1+M_2)\*(M_1^\beta+M_2^\beta)$ is closer to the
self-similar form for lower~$\beta$. The mass function for
$U\propto(M_1+M_2)^2$ is not self-similar (nonlocal case, see
discussion in Kontorovich et~al.\ \cite{physd}). However, for
$t>t_{\rm cr}$ the shape of the curve becomes nearly constant
(Fig.~\ref{fig4}a): a different self-similarity appears, because
mergers with the cD-galaxy dominate and the dependence of $U$ on
the smaller mass vanishes.

Both in direct solution of the Smoluchowski equation and in
numerical simulation of mergers, there exists a finite limit
mass: the mass of the sink in the former case, the total mass of
the system in the latter one. However, the problems which are
solved in this section and in Sect.~\ref{sec2} are not equivalent.
Nevertheless, simulation shows that, in spite of the essential
influence of the limit mass ($u_2>1$), the mass function obtained
by simulation of mergers in Sect.~\ref{sec2} and the one obtained
as a direct solution of the Smoluchowski equation have good
agreement in the region $M\ll M_{\rm max}$. So we can make a
conjecture that $\Phi(M)$ for $M\ll M_{\rm max}$ depends only on
the value of the limit mass and does not depend on its nature.

\section{Discussion}
\label{sec4}
Simulation confirms the possibility that massive galaxies may form
by mergers, moreover, this process has an ``explosive'' character
and is an analog of phase transition, cD-galaxies (with mass
comparable to the total galaxy mass of the cluster) being formed
as a new phase. What are the conditions which make this process
possible? The expression for $t_{\rm cr}$ may be written as
\begin{equation}
t_{\rm cr}=\frac{\xi_u}{c_uN_0M_0^u}
\label{25}
\end{equation}
(see, e.g., Voloshchuk \cite{vol}).
Numerical solution of the Smoluchowski equation gives
$\xi_2\approx0.02$, $\xi_{4/3}\approx0.26$, $\xi_{3/2}\approx0.1$
for $U=c_2(M_1+M_2)^2$, $U=c_{4/3}(M_1+M_2)\*(M_1^{1/3}+M_2^{1/3})$,
$U=c_{3/2}(M_1+M_2)\*(M_1^{1/2}+M_2^{1/2})$ respectively
(if the initial distribution $\Phi_0$ has a tail, $\xi_u$ may essentially
depend on $\Phi_0$ and be much less).
Assuming
$N_0\sim{\cal M}/M_0$ where ${\cal M}$ is an average density of
the mass contained in galaxies and expressing the variables in
astronomical units, we obtain the order of magnitude for~$t_{\rm
cr}$:
\begin{equation}
t_{\rm cr}\sim\cases{
2{\cdot}10^{15}\,v_7^3M_6^{-1}({\cal M}/\rho)^{-1}\mbox{ years}
&($u=2$),
\cr
4{\cdot}10^{13}\,v_7M_6^{-1/3}({\cal M}/\rho)^{-1}\mbox{ years}
&($u=4/3$),
\cr
10^{14}\,v_7M_6^{-1/2}({\cal M}/\rho)^{-1}\mbox{ years}
&($u=3/2$).
}
\label{26}
\end{equation}
Here $v_7=v_{\rm rms}/(10^7{\rm~cm~s^{-1}})$,
$M_6=M_0/(10^6~M_\odot)$, ${\cal M}/\rho$ is the ratio of the
local density of the mass contained in galaxies to the average
density of the Universe; the coefficient $C$ in Eq.~(\ref{6}) is
assumed $C\sim\frac{20{\rm~kpc}}
{(2{\cdot}10^{11}~M_\odot)^\beta}$. The
mass of a rich cluster is
$5{\cdot}10^{14}\mbox{--}5{\cdot}10^{15}~M_\odot$, 2--7\% of it is
contained in galaxies (B\"ohringer \cite{texas}).  The size (of the
central part) being of the order of one megaparsec, the ratio
${\cal M}/\rho$ may be several hundred to several thousand.
Assuming ${\cal M}/\rho\sim10^3$ we obtain that, for a cluster
with a low velocity dispersion ($\sim300{\rm~km~s^{-1}}$), the
critical time is less than the age of the Universe on condition
that masses of the initial galaxies (which then merge)
$M_0\sim10^9\mbox{--}10^{10}~M_\odot$ or more (for the case $M\gg
M_{\rm b}$, $u=3/2$). A close estimate for $M_0$ can be obtained
also for the region\footnote{The choice of the region depends on
the relation between $M_0$ and~$M_{\rm b}$. If $M_0\ll M_{\rm b}$
then the case $u=2$ is realized; if $M_0\ga M_{\rm b}$ then
$u=1+\beta$.} $M\ll M_{\rm b}$, $u=2$.  On the other hand, we may
consider the formation of massive galaxies and the mass function
tail only if the initial mass $M_0$ is much less than a typical
mass of a large galaxy ($\sim10^{11}~M_\odot$).
For a cluster with a bigger velocity dispersion
($1000{\rm~km~s^{-1}}$ or more) it is much more difficult to satisfy
condition~(\ref{26}): $M_{\rm b}$ is large, so the kernel with
$u=2$ should be taken; $t_{\rm cr}$ is proportional to $v_{\rm rms}^3$
and can be less than the age of the Universe only for very high density
(${\cal M}/\rho\sim10^4$) and large enough initial masses
($M_0\sim10^{10}~M_\odot$).
Thus, possible dependence of $v_{\rm rms}$ on time due to
cluster evolution, its space nonhomogeneity etc.\ can
essentially influence the role of mergers, especially on small
masses.

The estimate for ${\cal M}/\rho$ given above is based on the
assumption that dark matter belongs to the whole cluster rather
than to individual galaxies. If dark matter is concentrated in
galaxies\footnote{At least cD-galaxies may contain a large
quantity of dark matter (Ikebe et~al.\  \cite{ikebe}).}, the ratio
${\cal M}/\rho$ may increase by an order which results in the same
decrease of $t_{\rm cr}$ (according to Eq.~(\ref{26})).

So, the conditions necessary for the ``explosive'' process of
mergers may be realized in many clusters.

After a cD-galaxy which contain a significant part of the total
mass has formed in the cluster centre, the dynamics of the cluster
is largely determined by attraction to this galaxy, and the model
considered in this paper breaks down. Besides random collisions,
spatial inhomogeneity and mass segregation become essential: due
to dynamical friction, most massive galaxies gradually gets into
the centre and are swallowed by the cD-galaxy. However,
before~$t_{\rm cr}$, when there is no yet cD-galaxy in the
cluster, galaxy mergers can be considered as random pairwise
encounters with probability given by Eq.~(\ref{7}).
It is also clear that mass segregation
at a late stage of cluster evolution
should be computed together
with mass function evolution, using a spatially inhomogeneous
Smoluchowski type kinetic equation, which is a much more
complicated problem.

Note that galactic mergers due to dynamical friction were
discussed by Hausman \& Ostriker (\cite{ostriker}). In
spite of the difference in the model, the algorithm for simulation
of mergers, used in their work, is analogous to the algorithm we
use (and, thus, equivalent to the Smoluchowski equation), though
the expression for the merger probability is different ($U\propto
M_1M_2$, which also gives the ``explosive'' evolution).

As shown in Sects.~\ref{sec2}--\ref{sec3} (see also Kontorovich
et~al.\  \cite{physd}; Krivitsky \cite{jphys}), the mass function
formed by mergers with the probability given by Eq.~(\ref{7}) is
rather steep ($\alpha\approx2$ for
$U\propto(M_1+M_2)\*(M_1^\beta+M_2^\beta)$, $\alpha\sim2\mbox{--}3$
for $U\propto(M_1+M_2)^2$; in the latter case the asymptotics seem
to be non-power, so $\alpha$ cannot be determined accurately). It
is quite possible that the obtained values of the slope correspond
to the steepening of the cluster galaxy luminosity function at the
faint end, which was recently discovered (de~Propris et~al.\
\cite{deprop}; Kashikawa et~al.\  \cite{kashik}; Bernstein et~al.\
\cite{bernst}).  According to observational data analysed in these
works, the effective value of the slope for faint galaxies in
clusters increases up to 2--2.2 (though, as Bernstein et~al.\
note, there is a different interpretation of these results).
Possibly, this part of the luminosity function is formed by
mergers and can be described by the intermediate asymptotics for
the Smoluchowski equation
(if the latter can be extended to small enough masses).
Recent HST data shows also an excess of
faint low-mass objects for field galaxies (Cowie et~al.\
\cite{cowie}).

The appearance of relatively steep intermediate asymptotics
($\alpha\approx2$) can be easily understood from the following
arguments\footnote{We consider here the case
$U\propto(M_1+M_2)\*(M_1^\beta+M_2^\beta)$, when the mass
function is close to a power law.}.  Both obtained values for the
index ($\alpha\approx1.9$ for $u=4/3$ and $\alpha\approx2.1$ for
$u=3/2$) are within the range $(u+2)/2$ to $(u+3)/2$. The mass
function with $\alpha=(u+3)/2$ corresponds to a constant mass
flux\footnote{Solutions with a constant flux of a conserved
variable are analogous to Kolmogorov spectra in the weak
turbulence theory, see Zakharov et~al.\ \cite{turb}.} to infinity
(i.e., to cD-galaxy, in our case).  However, due to nonlocality of
the distribution\footnote{Nonlocality corresponds to divergence of
the collision integral for the power-law distribution.}
($\left|u_2-u_1\right|>1$, see Vinokurov \& Kats \cite{vk}) such a
solution is not realized exactly in both our cases. Nonlocality
leads to an essential role of interactions between low-mass and
high-mass galaxies. Then the number of massive galaxies is
approximately conserved, and the constant flux of their number to
infinity corresponds to $(u+2)/2$ (Kontorovich et~al.\ \cite{aat93}).
Since none
of these limit cases is realized, the index is situated between
these values:  $1.67<\alpha\approx1.9<2.17$ ($u=4/3$),
$1.75<\alpha\approx2.1<2.25$ ($u=3/2$), and is rather close to
their arithmetic mean value (as we can see both from the
simulation and the numerical solution of the Smoluchowski
equation).

The density ratio ${\cal M}/\rho$ in the above estimates was one
of the most important parameters which control the possibility of
effective merger process. The local value of this parameter may
vary in a very wide range: from~1 (scales exceeding an average
distance between massive field galaxies) to $10^7$ (if we take an
average density of a galaxy $\sim10^{-22}{\rm~g~cm^{-3}}$
for~${\cal M}$). As was shown above, in clusters this parameter is
large enough ($\sim10^3$ or even $10^4$) to yield an ``explosive''
evolution due to mergers. Local concentrations may enable
analogous phenomena for field galaxies at large~$z$ (see, e.g.,
Komberg \& Lukash \cite{lk}; Kontorovich \cite{obzor}).

Morphological changes in cluster galaxies, which is one of the
results of mergers, may be related to the change of the angular
momentum distribution (cf.~Toomre \cite{toomre}).
The possibility of dependence between Hubble's morphological type
and an effective angular momentum has been discussed in the
literature (see Fig.~1 in Polyachenko et~al.\ \cite{polyachenko})
and confirmed by an independent analysis of observational data
(Kontorovich \& Khodjachikh \cite{konkhod}; Kontorovich et~al.\
\cite{kkgb}). However, this dependence needs special consideration
which is beyond the scope of this work.

The above consideration of cluster evolution takes into account
only galaxy mergers in a spatially homogeneous model. It allows to
obtain the ``explosive'' evolution, the steep part of the mass
function, cD-galaxies, rapid evolution of galaxy morphological
types, and a mean value of the dimensionless angular momentum
which does not depend on the details of the initial distribution.
In the same time, this approach has obvious limitations. The
``explosive'' evolution does not produce Schechter's mass function
with $\alpha\approx1.25$. It is possible that the effective merger
probability $U$ changes at a late stage of cluster evolution (when
massive galaxies have formed) in such a way that $u_{\rm eff}$
becomes less than one and the ``explosive'' process slows
down\footnote{See examples of such $U$ changes in Kontorovich
et~al.\ \cite{ostanovka}; van~Dongen \cite{ucd}.}, which leads to
a flatter $\Phi(M)$. Another possibility is that this part of the
mass function may not be formed only by mergers.

\begin{acknowledgements}
This work was supported, in part, by the
International Soros Science Education Program through grants
PSU 052072 and SPU 042029.
\end{acknowledgements}

\end{document}